\documentclass[10pt]{article}

\usepackage{fancyhdr}
\usepackage{fullpage}
\usepackage{amsmath}
\usepackage{amsbsy}
\usepackage{amssymb}
\usepackage{amscd}
\usepackage{amsfonts}
\usepackage{graphics}
\usepackage{verbatim}
\usepackage{epsfig}
\usepackage{xspace}
\usepackage{euscript}
\usepackage{alltt}
\usepackage{multirow}
\usepackage{float}
\usepackage[colorlinks]{hyperref}
\usepackage{color}
\usepackage{t1enc}
\usepackage{times,exscale}
\usepackage{graphicx,calc}
\usepackage{subfig}
\usepackage{epstopdf}
\usepackage{verbatim}
\usepackage{geometry}
\usepackage{bm}
\def\be{{\bm{\epsilon}}}
\def\bsigma{{\bm{\sigma}}}

\title{\emph{Influence of thermomechanical loads on the energetics of precipitation in magnesium aluminum alloys}}
\date{\vspace{-10ex}}

\begin{document}
\maketitle
\begin{center}

{{Swarnava Ghosh and Kaushik Bhattacharya\\ Division of Engineering and Applied Science \\ California Institute of Technology}}
\end{center}
\begin{abstract}
We use first principles calculations to study the influence of thermomechanical loads on the energetics of precipitation in magnesium-aluminum alloys. Using Density Functional Theory simulations, we present expressions of the energy of magnesium-aluminum binary solid solutions as a function of concentration, strain and temperature. Additionally, from these calculations, we observe an increase in equilibrium volume (and hence the equilibrium lattice constants) with the increase in concentration of magnesium. We also observe an increase in the cohesive energy of solutions with increase in concentration, and also present their dependence on strain. Calculations also show that the formation enthalpy of $\beta$ phase solutions to be strongly influenced by hydrostatic stress, however the formation enthalpy of $\alpha$ phase solutions remain unaffected by hydrostatic stress. We present an expression of the free energy of any magnesium aluminum solid solution, that takes into account the contributions of strain and temperature. We note that these expressions can serve as input to process models of magnesium-aluminum alloys. We use these expressions to report the influence of strains and temperature on the solubility limits and equilibrium chemical potential in Mg-Al alloys. Finally, we report the influence of thermomechanical loads on the growth of precipitates, where we observe compressive strains along the $c$ axis promotes growth, whereas strains along the $a$ and $b$ directions do not influence the growth of precipitates. 

\end{abstract}
\section{Introduction}

Magnesium is the lightest among all commonly used structural metals, abundantly available on the earth's crust and has among the highest strength to weight ratio.  Further, the availability of high purity magnesium, low melting temperature and high specific heat offer attractive potential for a wide range of applications including automotive, aeronautics and aerospace to consumer electronics.
Still, the lack of ductility and fracture toughness has limited the actual applications despite considerable research over the last decade \cite{Kainer:2000,Nie:2012,Polmear:1994}. 

Magnesium has a hexagonal closed-packed (HCP) structure with very easy slip along the basal plane and difficult slip along the pyramidal plane.  This high anisotropy contributes to the strength with proper texture but also to the low ductility.  Aluminum is a commonly used alloying element -- including in the AZ class of magnesium alloys --  because of its availability, low density, desirable corrosion and mechanical properties.  Aluminum forms $\beta$ phase precipitates, an intermetallic compound with a composition of Mg$_{17}$Al$_{12}$ and a body-centred cubic structure (space group $I\bar{4}3m$).  These precipitates strengthen magnesium alloys by resisting the motion of basal dislocations \cite{Kainer:2000,Nie:2012,Polmear:1994,Nembach:1997}.  Significant efforts have been directed towards characterizing and understanding precipitation strengthening in magnesium alloys \cite{Wilson:2003,He:2006,Buha:2008,Zhang:2018,Nakata:2017}.  It has been recognized that the amount of hardening depends critically on the number density and orientation \cite{Nie:2003}. Further, while the intermetallic precipitates contribute to hardening, they also can contribute to the lack of ductility, fracture and fatigue, and reduced spall strength when their sizes become too large \cite{Ma:2019,Prameela:2019,Lloyd:2019}.  Therefore, there is an interest in controlling the size, orientation and number density of precipitates by thermo-mechanical treatment.

The conventional processing route involves solution treatment at ~692 K within a single phase region of magnesium with an HCP structure, followed by quenching to a supersaturated solid solution of aluminum in magnesium matrix, and finally ageing at a temperature range of 373 K to 573 K, which decomposes the solid solution into a fine distribution of precipitates in the magnesium matrix. There are two ways of formation of $\beta$ phase precipitates. First is by discontinuous precipitation, which give rise to a lamellar structure. Discontinuous precipitates start growing from the grain boundaries and expands towards the grain center\cite{Nie:2012}, and develop primarily at temperatures of up to 450 K \cite{Braszczynska:2009}. The second type of precipitates is continuous precipitation which forms in the grain interiors, have smaller aspect ratios than discontinuous precipitates, and typically develops during annealing of a super saturated alloy at temperatures greater than 600K \cite{Braszczynska:2009}. In AZ class of magnesium alloys, both continuous and discontinuous precipitation occur simultaneously at intermediate temperatures\cite{Braszczynska:2009, Duly:1995, Nie:2012}. Heat treatment conditions also affect the type of precipitation \cite{Avedesian:1999,Caceres:2002,Maitrejean:1999,Nie:2001}. 

Certain conditions of thermomechanical processing involving deformation and temperature not only recrystallizes grain, but can also drive precipitation \cite{Park:2012,Zhang:2011,Mathis:2005,Ma:2019}. Unlike the precipitates formed by conventional ageing, precipitates formed during dynamic precipitation are continuous, nanosized with high number density and low aspect ratio \cite{Mathis:2005,Ma:2019}.  While this is encouraging, there is a  lack of complete understanding which can accurately capture deformation driven precipitation, and this provides the motivation for the current work.

We use density functional theory to calculate the energies of magnesium-aluminum phases under diverse conditions of temperature and strain, and use it to find the influence of thermomechanical loads on the energetics of precipitation in magnesium-aluminum binary alloys.  This provides insights into precipitation, and also free energy surfaces that can be inputs to process models.  

The remainder of this manuscript is organized as follows. In Section \ref{Sec:CompApproach} we describe the details of the first principles calculations.  We present and discuss the results of our simulations in Section \ref{Sec:Results} . Finally, in Section \ref{Sec:Conclusion}, we provide concluding remarks.

\section{Computational approach}\label{Sec:CompApproach}

We use the first principles density functional theory in a real space framework using the SPARC \cite{Ghosh:2017a,Ghosh:2017b} platform.  We  employ a twelfth-order accurate finite-difference discretization with a mesh spacing of 0.5 Bohr and the Perdew-Wang parameterization \cite{Perdew:1992} of  the correlation energy calculated by Ceperley-Alder \cite{Ceperley:1980} as well as norm-conserving Troullier-Martins pseudopotentials \cite{Troullier:1991}. The SPARC framework uses polynomial filtered subspace iteration associated with Chebyshev polynomials of degree 20.  We assume periodicity and use  a $10 \times 10\times 10$ Monkhorst-Pack grid for performing all integrations over the Brillouin zone.  Finally, we accelerate the convergence of the self consistent field iteration using periodic Pulay mixing \cite{Banerjee:2016} with mixing history 7, mixing parameter 0.2 and Pulay extrapolation frequency 2.

\begin{figure}
\centering
\includegraphics[keepaspectratio=true,width=0.4\textwidth]{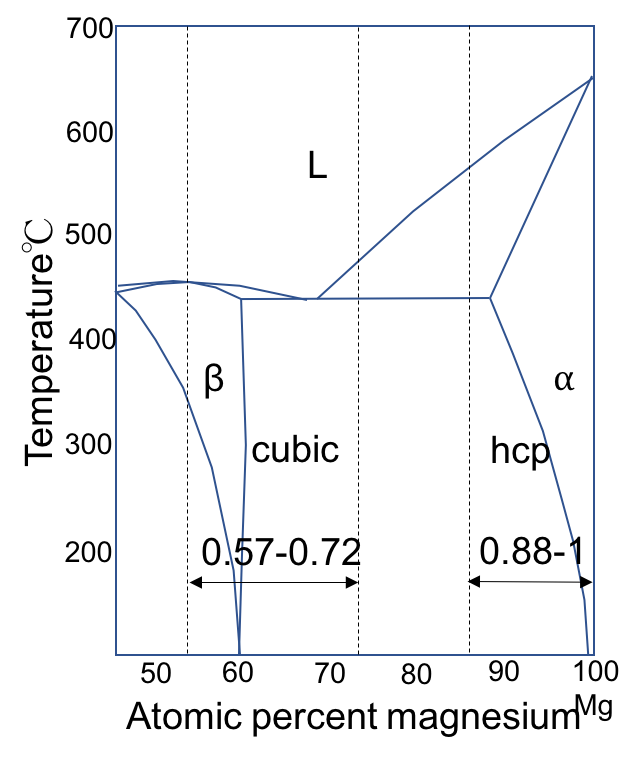}
\caption{The partial phase diagram of the Aluminum-Magnesium binary system, and the regions studied.}
\label{Fig:Schematic}
\end{figure}

We consider two phases of Mg-Al marked in the partial phase diagram reproduced in Figure \ref{Fig:Schematic}.  The first is the $\alpha$ phase that is a hexagonal close packed Mg phase with Al as solute.  We study this phase with Mg concentration ranging from 0.875 to 1.0. We use a 4 atom unit cell for a pure Mg phase. For the other Mg rich phases, the number of atoms in the supercells are 8, 16 and 32 for concentrations of 0.875, 0.938 and 0.969, respectively. In these supercells, one Mg atom is substituted with an Al atom. In each of these supercells, there are multiple lattice sites at which the Al substitution can occur. In our work, we have calculated the energies of a few of these cases, and have observed that the variation in energy between different substitutions is less than $4 \times 10^{-4}$ eV/atom, which is much less than the variation with respect to concentration, and strain.  The second phase is the $\beta$ phase that is a cubic Mg$_{17}$Al$_{12}$ phase with either Al or Mg solutes.  We study this phase with concentration ranging from  0.569 to 0.724,  and 58 atoms in the unit cell. 

For each configuration, we first find the ground state by letting the atoms relax under the constraint of crystallographic symmetry till the  maximum component of atomic force is less $1 \times 10^{-4}$ Ha/Bohr.  To study the effect of strain, we apply the strain relative to this ground structure on the outer atoms of the computational cell and let the internal atoms relax till the  maximum component of atomic force is less $1 \times 10^{-4}$ Ha/Bohr.   We have nine realizations of strain at each concentration.

\section{Results and discussion}\label{Sec:Results}

\subsection{Equilibrium volume and lattice constant}\label{Subsection:EQ}

\begin{figure}\center
\subfloat[equilibrium volume and lattice constant]{\includegraphics[keepaspectratio=true,width=0.8\textwidth]{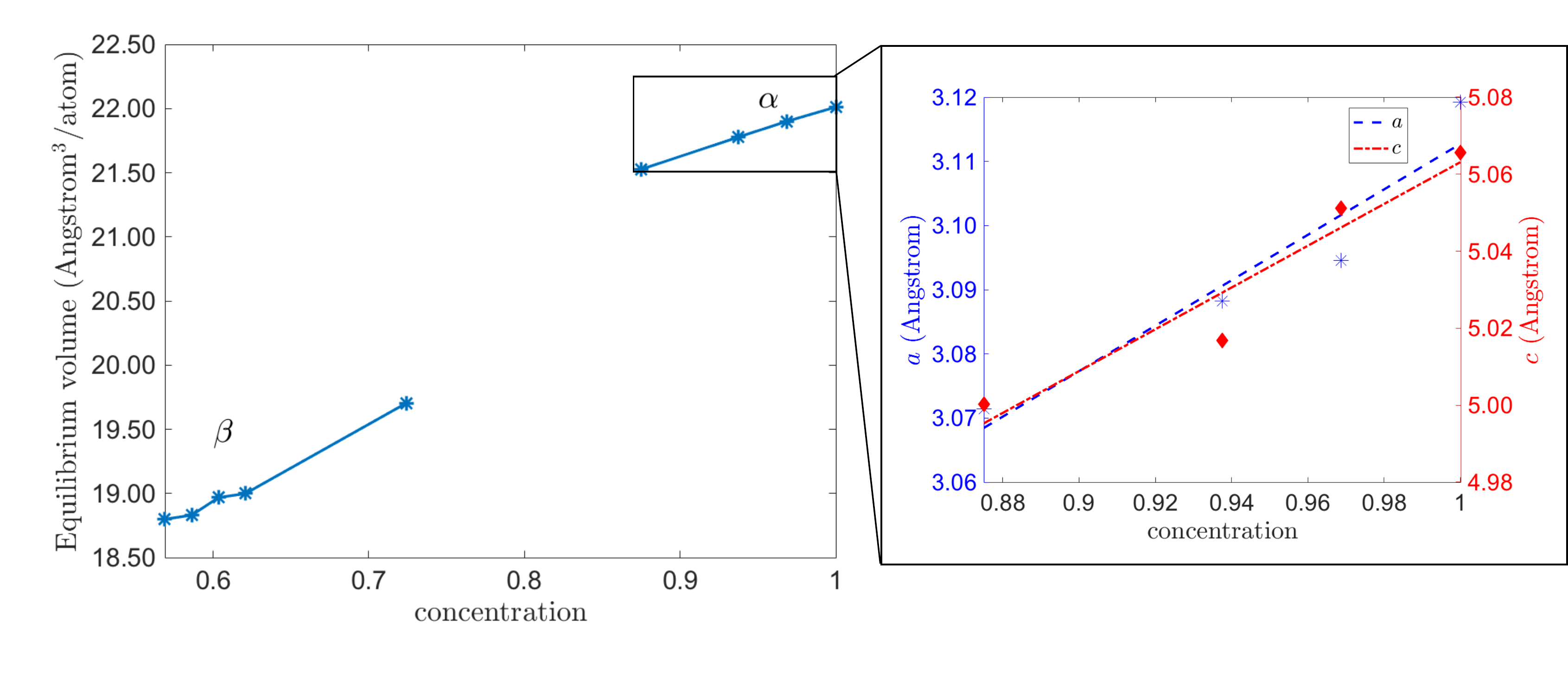}\label{Fig:Volume}}\\
\subfloat[electron density contours]{\includegraphics[keepaspectratio=true,width=0.9\textwidth]{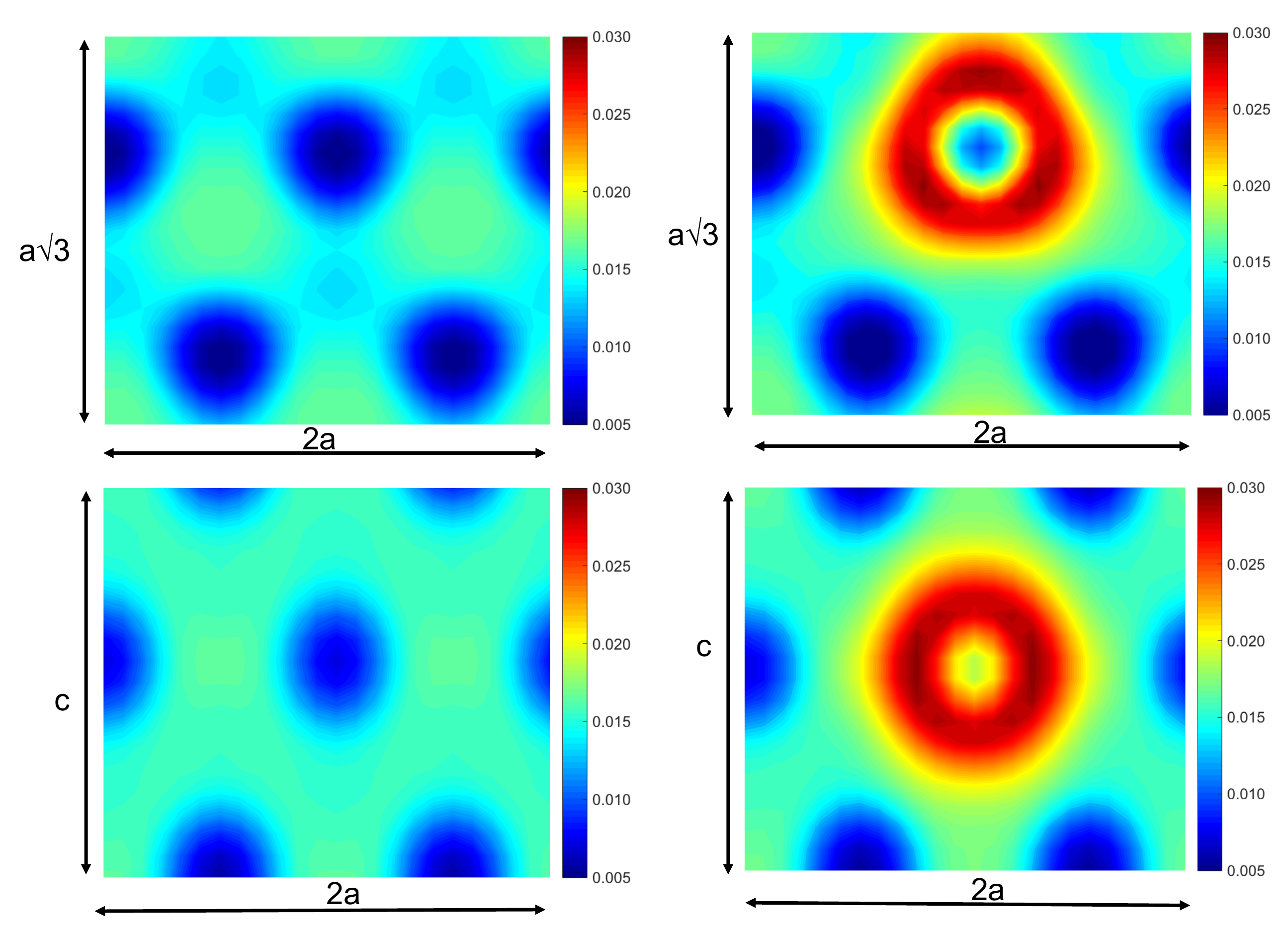}\label{Fig:Electrondensity}}
\caption{ a) Equilibrium volume and lattice parameters as a function of concentration and Dependence of the lattice constants of binary sold solutions with pure magnesium ($\alpha$ phase) as the solvent on the concentration of magnesium in the solution. (b) Top panel shows the electron density on the basal plane of lattice of pure Mg (left) and that in the vicinity of Al atom in an Mg lattice with dissolved Al (right). Bottom panel shows the electron density on the $c-a$ plane of lattice of pure Mg (left) and that in the vicinity of Al atom in an Mg lattice with dissolved Al (right).} 
\label{fig:volden}
\end{figure}

The equilibrium lattice constants are calculated to be $a$=3.125 Angstrom, $c/a$=1.624 for pure magnesium (pure $\alpha$) and $a$=10.301 Angstrom for Mg$_{17}$Al$_{12}$ (pure $\beta$) . Previously reported values of the lattice constant of pure magnesium are $a$=3.16 Angstrom, $c/a$=1.61 (DFT) \cite{Chou:1986}, $a$=3.21 Angstrom, $c/a$=1.623 (experiment)\cite{Kittel,Koster:1961}, and the equilibrium lattice constant of the $\beta$ phase is 10.56 Angstrom (DFT) \cite{Duan:2011}, and 10.40 Angstrom (experiment) \cite{Zhang:2005}.  Figure \ref{Fig:Volume} shows that the equilibrium volume decreases with decreasing concentration  of magnesium in both phases.  In fact, both  $c$ and $a$ decrease with decreasing concentration  of magnesium in the $\alpha$ phase as also shown in agreement with previously reported trend\cite{Shin:2010}. The change in the concentration of magnesium in the $\alpha$ phase solutions has negligible influence on the $c/a$ ratio. We fit the equilibrium volume as a function of concentration: 
\begin{equation}
V^{\alpha,\beta}[c] = V_1^{\alpha,\beta} (c-c^{\alpha,\beta}) + V_0^{\alpha,\beta}
\end{equation}
where $V_0^{\alpha,\beta}=\{22.02,18.85\}$ Angstrom$^3$/atom and $V_1^{\alpha,\beta}=\{3.904,6.029\}$ Angstrom$^3$/atom, and $c^\alpha = 1, c^\beta = 0.586$.

Figure  \ref{Fig:Electrondensity} compares the electron density on the basal plane in a lattice with pure Mg with that in a lattice with an Al solute.   We observe that there is an increased electron density in the vicinity of an Al atom compared to that of a Mg atom.  This reflects the higher valency of Al (3) compared to that of Mg (2).  This elevated electron density causes a tight bonding of Al to the neighbors compared to Mg resulting in a decrease in lattice parameters.

\subsection{Energy}

\begin{figure}
\subfloat[Energy]{\includegraphics[keepaspectratio=true,width=0.5\textwidth]{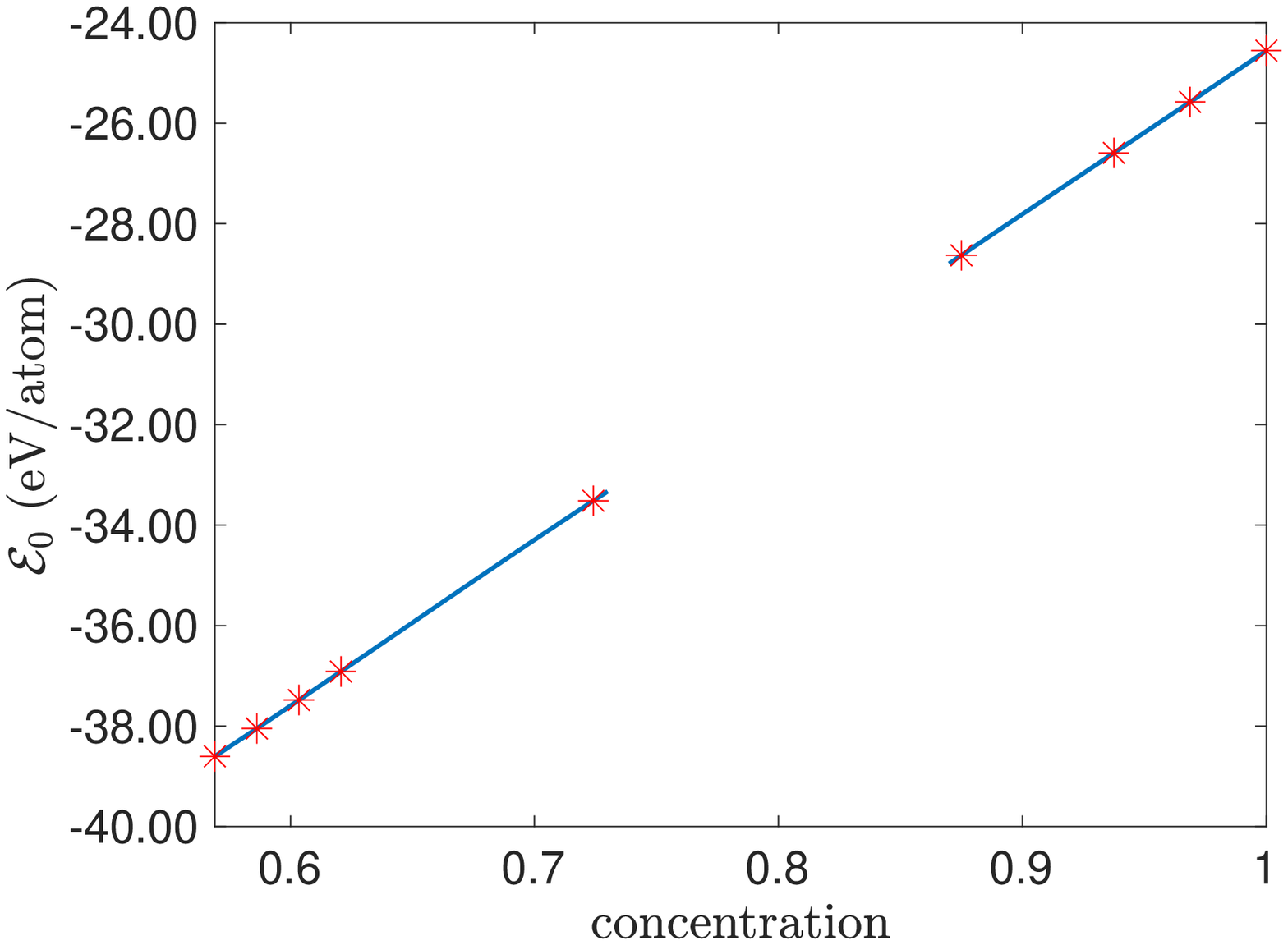}\label{Fig:EVsConc}}
\subfloat[Energy]{\includegraphics[keepaspectratio=true,width=0.5\textwidth]{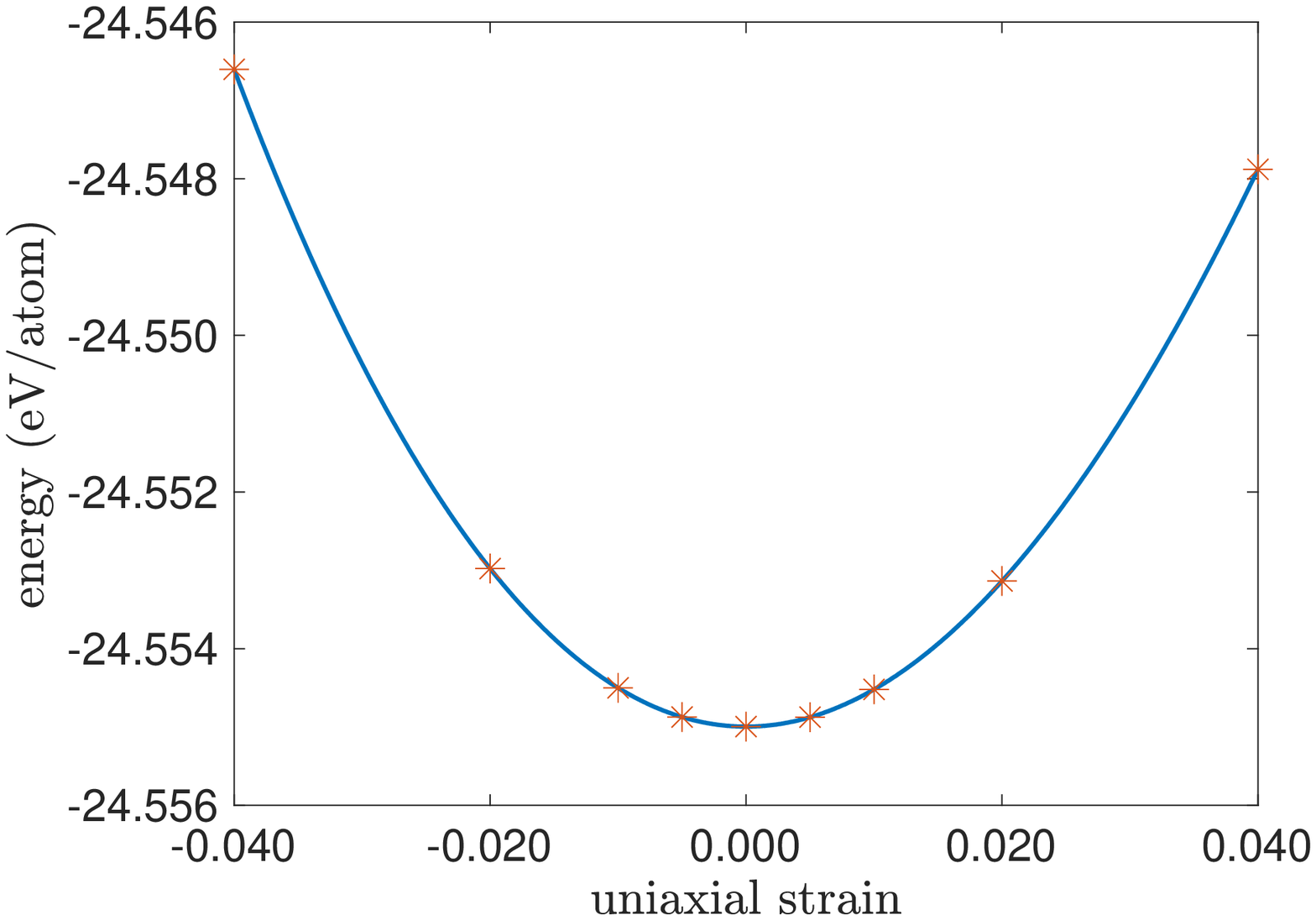}\label{Fig:aVsConc}}\caption{(a) Energy vs. concentration. (b) Energy vs. uniaxial strain along the $c$-axis for Pure Mg. Notice the tension-compression asymmetry of the energy of pure Mg. In (a) and (b), the stars are the computed points and the solid line is the fit. } 
\label{fig:energy}
\end{figure}

Next we turn to the computed (internal) energy at various concentrations and strains. See Supplementary information for the data. Figure \ref{fig:energy} shows representative results. We find it possible to collapse the results of all our computations for each phase into the expression
\begin{equation}\label{Eqn:EnergyFit}
\mathcal{E}^{\alpha,\beta}[c,\be] = {\mathbb M}^{\alpha,\beta}_{ijklmn} [c] (\epsilon_{ij}-\epsilon_{ij}^{\alpha,\beta}[c])(\epsilon_{kl}-\epsilon_{kl}^{\alpha,\beta}[c])(\epsilon_{mn}-\epsilon_{mn}^{\alpha,\beta}[c]) + {
\mathbb C}^{\alpha,\beta}_{ijkl} [c](\epsilon_{ij} - \epsilon_{ij}^{\alpha,\beta}[c])(\epsilon_{kl}-\epsilon_{kl}^{\alpha,\beta}[c])  + \mathcal{E}_0^{\alpha,\beta}[c] \,,
\end{equation}
where the phase and concentration dependent ground state (strain-free) energy $\mathcal{E}^{\alpha,\beta}_0$ as well as the phase and concentration dependent second and third order moduli ${\mathbb M}^{\alpha,\beta}$ and ${\mathbb C}^{\alpha,\beta}$ are 
\begin{eqnarray}
\mathcal{E}^{\alpha,\beta}_0[c] &=& E^{\alpha,\beta}_3 (c-c^{\alpha,\beta})^3 + E^{\alpha,\beta}_2 (c-c^{\alpha,\beta})^2 + E^{\alpha,\beta}_1 (c-c^{\alpha,\beta}) + E^{\alpha,\beta}_0, \\
{\mathbb M}^{\alpha,\beta}_{ijklmn} [c] &=& ({\mathbb M}_1)^{\alpha, \beta}_{ijklmn} (c-c^{\alpha,\beta}) + ({\mathbb M}_0)^{\alpha, \beta}_{ijklmn}, \\
{\mathbb C}^{\alpha,\beta}_{ijkl} [c] &=&  ({\mathbb C}_2)^{\alpha, \beta}_{ijkl} (c-c^{\alpha,\beta})^2 +  ({\mathbb C}_1)^{\alpha, \beta}_{ijkl} (c-c^{\alpha,\beta}) + ({\mathbb C}_0)^{\alpha, \beta}_{ijkl} .
\end{eqnarray}
The fitted coefficients (as well as the root-mean-square error) are shown in Table \ref{Table:Efit}.  $c^{\alpha,\beta}$ are concentrations of the pure $\alpha$ and $\beta$ phases ($c^\alpha = 1, c^\beta = 0.586$).  The strain $\epsilon_{ij}^{\alpha,\beta}[c]$ is the eigenstrain due to the expansion of the lattice with increase in magnesium concentration, and is measured with respect to the pure Mg or Mg$_{17}$Al$_{12}$. This can be calculated from the fits of the equilibrium volume given in Section \ref{Subsection:EQ}.

\begin{table}
\centering
{\bf Zero strain coefficients}\\ \vspace{0.1in}
\begin{tabular}{cccccc}
\hline
solvent phase &$E_0$&$E_1$&$E_2$&$E_3$ & RMSE \\   
\hline
$\beta$ & -38.0461 & 32.5114 & 12.1023 & -69.9654 & 0.002 \\
\hline  
$\alpha$ &  -24.5554 & 32.8235 & 6.1759 & 34.5826 & 0.002 \\
\hline
\end{tabular}
\vspace{0.2in}

{\bf Elastic constants}\\ \vspace{0.1in}
\begin{tabular}{cccccc}
\hline
solvent phase & strain & ${\mathbb C}^0_{ijkl}$ & ${\mathbb C}^1_{ijkl}$ & ${\mathbb C}^2_{ijkl}$ & RMSE\\   
\hline
\multirow{3}{*}{$\beta$} & volumetric & 2.914  & 39.49 & -255.2 & 0.258 \\
        & \{1111,2222,3333\} & 5.482 & 3.624 & -503.7 &  0.033 \\
        & \{1122,1133,2233\} & 1.630 & 57.423 & -130.950 & - \\
\hline  
\multirow{4}{*}{$\alpha$} & volumetric & 2.596 &  -7.676 & -66.41 & 0.020\\
         & \{1111\}   & 5.017 & -13.61 & -92.49 & 0.214 \\
         & \{2222\}   & 4.473 & -12.39 & -184.1 & 0.186\\
         & \{3333\}   & 4.847 & -5.27 & 36.86 & 0.152 \\
\hline
\end{tabular}
\vspace{0.2in}

{\bf Third-order Elastic constants}\\ \vspace{0.1in}
\centering
\begin{tabular}{ccccc}
\hline
solvent phase & strain &${\mathbb M}^0_{ijklmn}$&${\mathbb M}^1_{ijklmn}$ & RMSE \\   
\hline
\multirow{2}{*}{$\beta$} & volumetric & 0 & 0 & - \\
        & \{111111,222222,333333\} & 0 & 0 & - \\
\hline  
\multirow{4}{*}{$\alpha$} & volumetric   & -3.951 & 10.460 & 0.131 \\
 						  & \{111111\}   & -9.612 & 71.74 &  0.160 \\
                          & \{222222\}   & -11.95 & 6.326 &  0.152 \\
                          & \{333333\}   &  -9.961 & 0.018 &  0.135 \\
\hline
\end{tabular}
\caption{Fitted constants of energy surface (\ref{Eqn:EnergyFit}). All constants are in eV/atom}
\label{Table:Efit}
\end{table}

\subsection{Cohesive energy}
The relative stability of crystalline structure is determined from the cohesive energy defined as the energy difference between the crystallized structure and the isolated atoms.  The cohesive energy per atom is defined as
\begin{equation} \label{Eqn:CohesiveEnergy}
E^{\alpha,\beta}_{coh}[c;\be] = \mathcal{E}^{\alpha,\beta}[c;\be] - c \,\mathcal{E}^{Mg} - (1-c)\mathcal{E}^{Al} \,\,,
\end{equation}
where $\mathcal{E}^{Mg}$ and $\mathcal{E}^{Al}$ are the energies of isolated magnesium and aluminum atoms, respectively.  Note that a negative value implies that the crystallized form is more stable than the isolated atoms and lower value implies greater stability of the crystallized form.

\begin{figure} 
\centering
\includegraphics[keepaspectratio=true,width=0.50\textwidth]{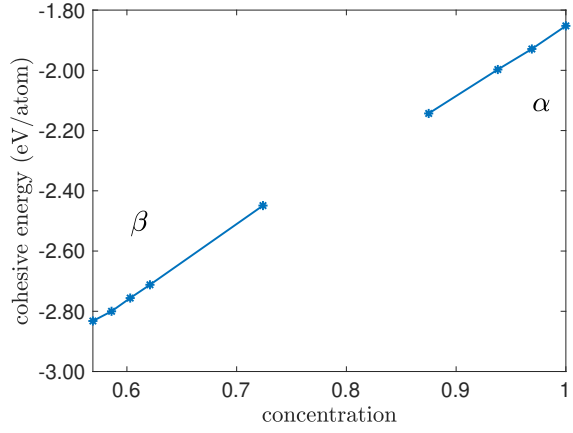}
\caption{ Cohesive energy vs. concentration of magnesium for the two  phases.} \label{fig:ceconc}
\end{figure}

We find that the cohesive energy of the $\beta$ phase is -2.80 eV/atom and the cohesive energy of the $\alpha$ phase is -1.89 eV/atom. Previously reported values of the cohesive energy of $\beta$ phase is -2.56 eV/atom (DFT) \cite{Duan:2011}, and for the $\alpha$ phase is -1.64 eV/atom (DFT-LDA)\cite{Chou:1986}, and  -1.51 eV/atom (experiment) \cite{Kittel,Koster:1961}. Figure \ref{fig:ceconc} shows the cohesive energy of unstrained solid solutions as a function of concentration of magnesium. We observe that the calculated values of cohesive energy is negative for both types of solid solutions. Further, the cohesive energy increases with the increase in concentration of magnesium, signifying decreasing stability of solutions with increasing concentration.  This is again consistent with our observation that Al atoms bind more tightly with its neighbors compared to the Mg atoms (see discussion of Figure \ref{fig:volden} in Section \ref{Subsection:EQ})).

\begin{figure}
\centering
\subfloat[volumetric strain]{\includegraphics[keepaspectratio=true,width=0.5\textwidth]
{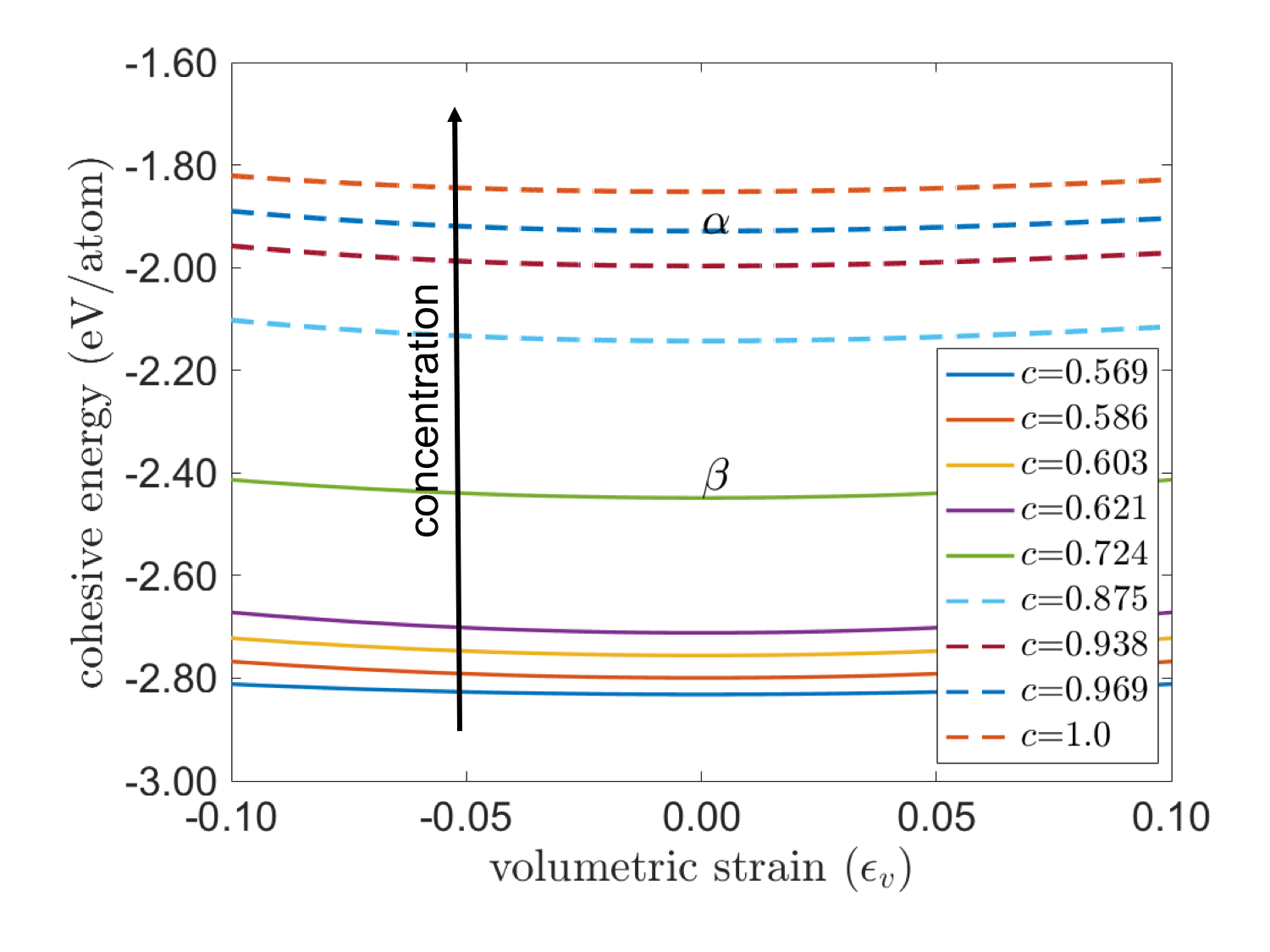}\label{Fig:CEStrain}}
\subfloat[uniaxial strain]{\includegraphics[keepaspectratio=true,width=0.5\textwidth]
{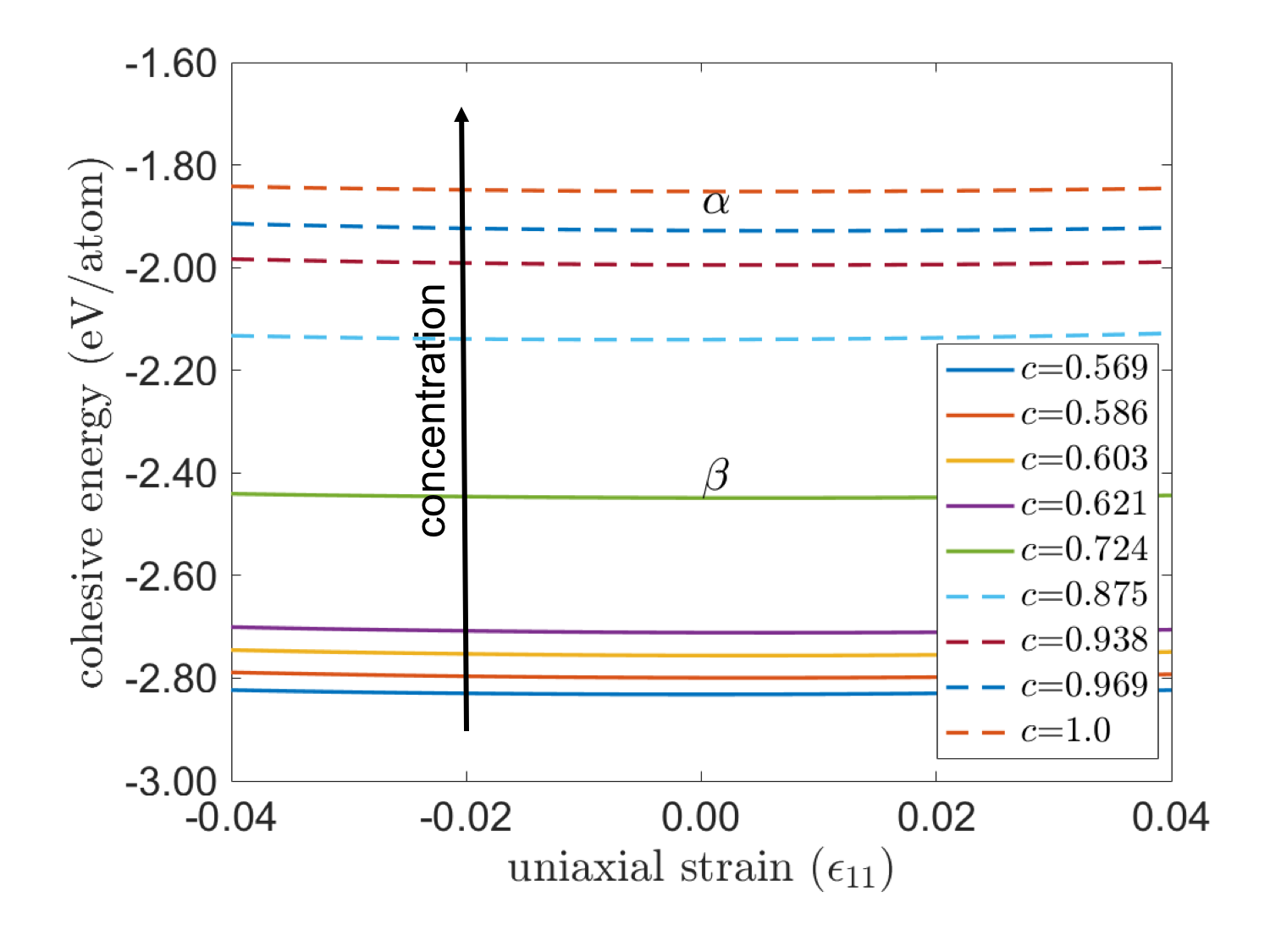}\label{Fig:CEStrainUniaxialc}}\\
\subfloat[volumetric strain]{\includegraphics[keepaspectratio=true,width=0.50\textwidth]{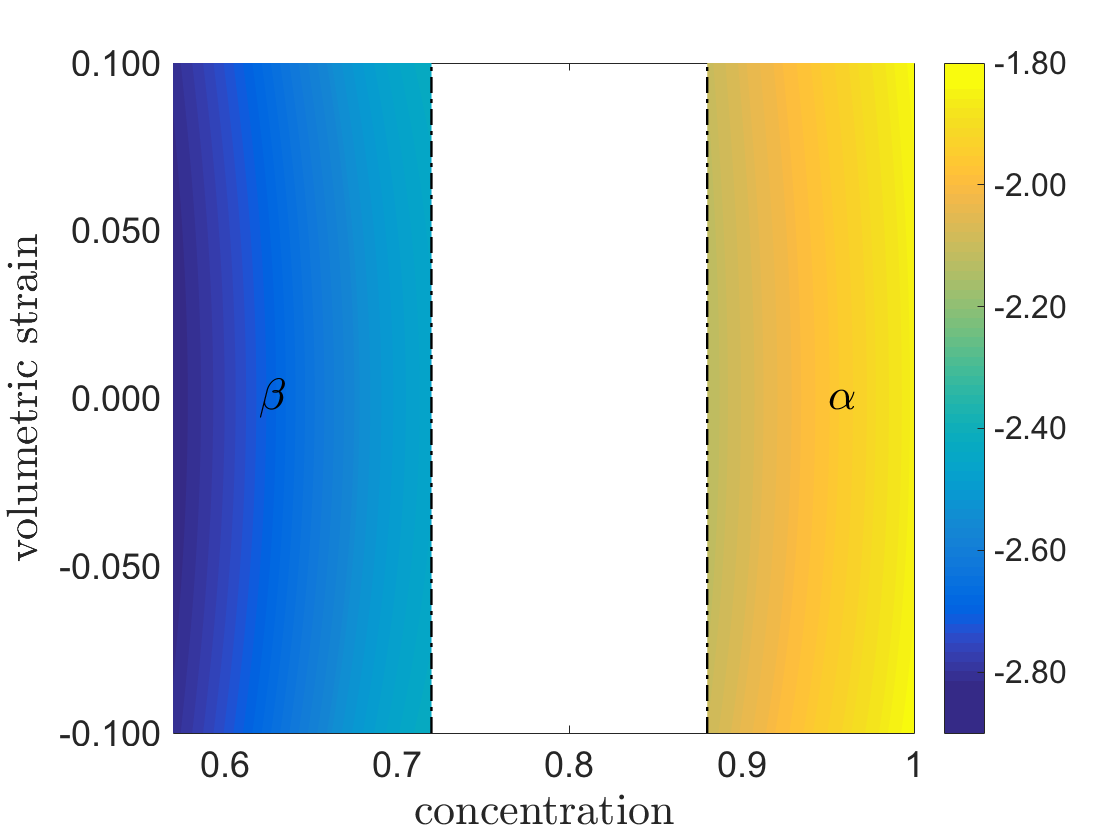}\label{Fig:CESurfaceVolumetric}} 
\subfloat[uniaxial strain]{\includegraphics[keepaspectratio=true,width=0.50\textwidth]{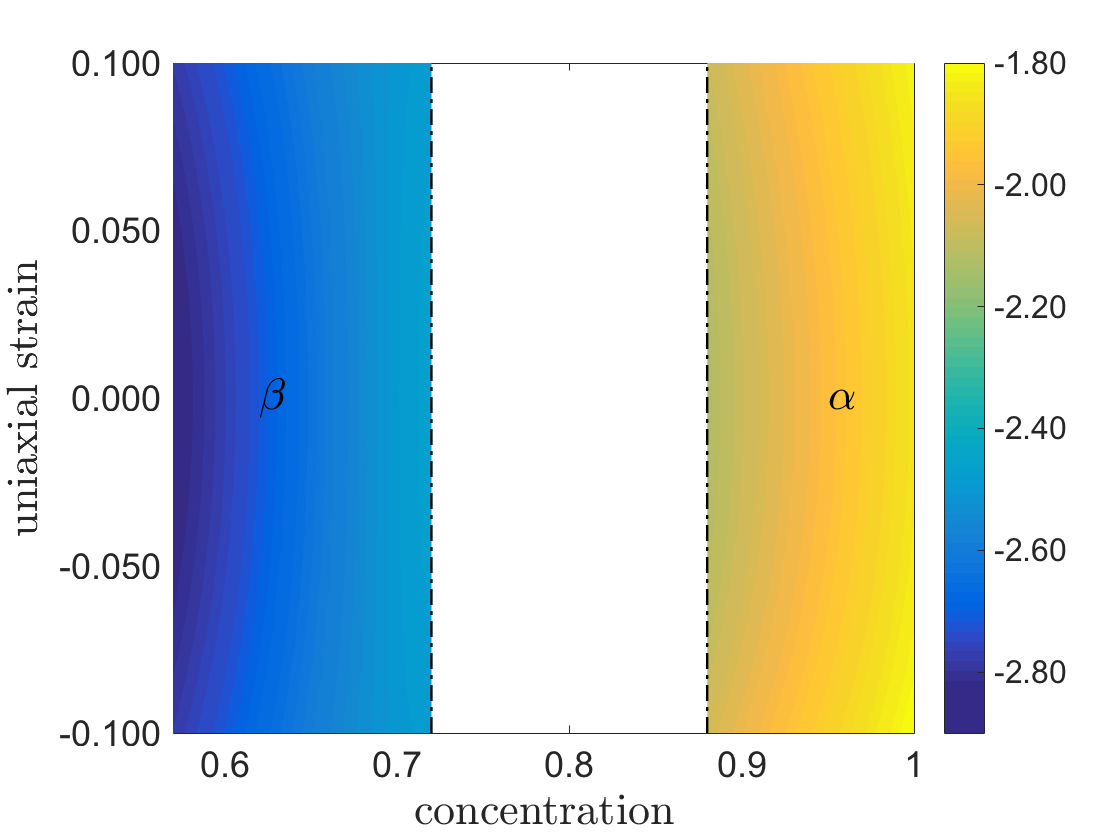}\label{Fig:CESurfaceUniaxial_c}} \\
\caption{Cohesive energy as a function of strain and concentration.  (a,c) Volumetric strain, (b,d) uniaxial strain along the cubic axis ($\beta$ phase) and $c-$ axis ($\alpha$ phase).}
\label{fig:ce}
\end{figure}

Figure \ref{fig:ce} shows how the cohesive energy depends on concentration and strain.  We display volumetric strain and uniaxial strain along the $c-$axis for the $\alpha$ phase and cubic axis for the $\beta$ phase.  The curvature with respect to the strain gives the moduli.  We can use the expressions (\ref{Eqn:EnergyFit}), and (\ref{Eqn:CohesiveEnergy}) to infer that these correspond to ${\mathbb C}^{\alpha,\beta}_{ijkl}$.  We find that the bulk modulus is 54 GPa for the $\beta$ phase and 37 GPa for pure magnesium. The previously reported values of the bulk modulus is 49.6 GPa (DFT) \cite{Wang:2008} for the $\beta$ phase and 35 GPa (DFT) \cite{Chou:1986}, 35.4 GPa (experiment) for pure magnesium.

\subsection{Formation enthalpy}

The enthalpy of the $\alpha$ and $\beta$ phases is the Legendre transform of the internal energy
\begin{equation}
\mathcal{H}^{\alpha,\beta}[c,{\mathbf \sigma}]=\min_{\be} \left\{ \mathcal{E}^{\alpha,\beta}[c,\be]+ (\be-\be^{\alpha,\beta}):{\mathbf \sigma} \right\} \,\,.
\end{equation}
The formation enthalpy $\Delta \mathcal{H}^{\alpha,\beta}$ is
\begin{equation}
\Delta \mathcal{H}^{\alpha,\beta}[c, {\mathbf \sigma}] = \mathcal{H}[c; \sigma]- c \,\mathcal{H}_{Mg}[\sigma]-(1-c) \mathcal{H}_{Al}[\sigma] \,\,, 
\end{equation}
where $\mathcal{H}^{\alpha,\beta}$ is the enthalpy of the solid solution, $\mathcal{H}^{Mg}$ is the enthalpy of pure magnesium and $\mathcal{H}^{Al}$ is the enthalpy of pure aluminum, all taken at  stress ${\mathbf \sigma}$.

We use formation enthalpy to compare the relative alloying ability of the solutions, with lower value suggesting stronger alloying ability \cite{Wang:2008}.

\begin{figure}
\centering
{\includegraphics[keepaspectratio=true,width=0.55\textwidth]{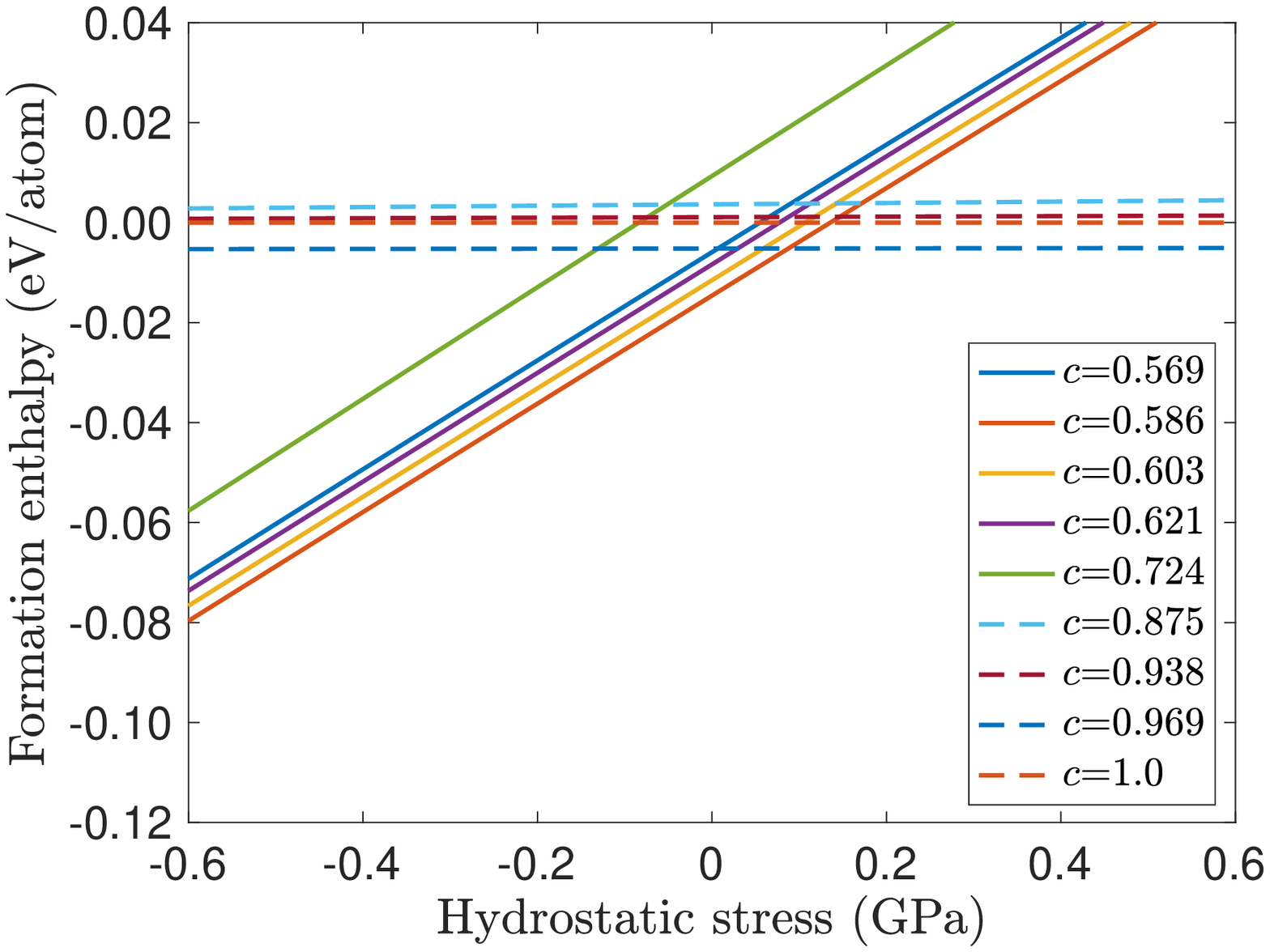}}
\caption{Effect of hydrostatic stress on the formation enthalpy. The solid curves denote solutions with $\beta$ phase as the solvent, the dashed curves denote solutions with $\alpha$ phase as solvent.}\label{Fig:EnthalpyStress}
\end{figure}

Figure \ref{Fig:EnthalpyStress} shows the dependence of the formation enthalpy on hydrostatic stress.   We note that it is common to encounter these stresses within the range from $\pm$ 0.6 GPa during thermomechanical processing of magnesium aluminum alloys. 
 From the figure, we see that for all the values of hydrostatic stress, the formation enthalpy of the $\beta$ phase is the lowest among all solid solutions considered in our work. Also evident from the slopes of the plots, hydrostatic stress has considerable influence on the formation enthalpy of the solutions with $\beta$ phase as solvent but has negligible effect on the formation enthalpy of the solutions with $\alpha$ phase as solvent. For the solutions with $\beta$ phase as solvent, tensile hydrostatic stresses further decrease the formation enthalpy, whereas compressive hydrostatic stresses increase the formation enthalpy. 

\subsection{Free energy}
The Helmholtz free energy as a function of concentration $c$, strain $\be$ and temperature $T$ is given by 
\begin{equation} \label{Eqn:FE}
\mathcal{F}^{\alpha,\beta}[c,\be,T] = \mathcal{E}^{\alpha,\beta}[c,\be] - T S^{\alpha,\beta}[c,\be] \,\,.
\end{equation}
In this work, we limit the entropy to the configurational entropy (i.e., we neglect the vibrational entropy {\footnote{Calculating vibrational entropy requires accurate calculation of phonon frequencies using Density Functional Perturbation Theory (DFPT). The cell sizes employed in this work are too large for currently available DFPT codes.}).  This is independent of phase and strain, and is given by 
\begin{equation}
S^{\alpha,\beta}[c,\be] = S[c] = -k_B (c \log c + (1-c)\log (1-c)) \,,
\end{equation}
where $k_B$ is the Boltzmann constant.   Recalling the expression (\ref{Eqn:EnergyFit}) for the internal energy, we have an expression for the free energy. This free energy functional can be employed for phase field simulations of deformation assisted precipitation in magnesium aluminum alloys.


\paragraph*{Growth of precipitates under thermomechanical loads} 
We now use the derived free energy expression to analyze the conditions of temperature and applied strain on the growth of precipitates. In Mg-Al alloys, the orientation relationship between the $\beta$ and $\alpha$ phase is $(0001)_{\alpha} = (110)_{\beta}$, $[1\bar{2}10]_{\alpha} \parallel [1\bar{1}1]_{\beta}$ \cite{Han:2015,Nie:2012}. The transformation strain associated with this hcp to bcc transformation is given by \cite{Han:2015}
\begin{equation}\label{Eqn:TS}
\epsilon_T^{\beta} = \begin{pmatrix}
\frac{a_{\beta}\sqrt{3} - 6a}{6a} & 0 & 0 \\
0 & \frac{a_{\beta}\sqrt{11} - 6\sqrt{3}a}{6\sqrt{3}a} & 0 \\
0 & 0 & \frac{a_{\beta}\sqrt{2} - 3c}{3c}
\end{pmatrix} 
=
\begin{pmatrix}
-0.0484 & 0 & 0 \\
0 & 0.0520 & 0 \\
0 & 0 & -0.0432
\end{pmatrix} 
\end{equation}
where $a_{\beta}$ is the lattice constant of the $\beta$ phase, and $a$, and $c$, are the lattice constants of the $\alpha$ phase. The transformation strain in the $\alpha$ phase is zero. The transformation strains are assumed to be independent of the change in concentration within each phase.
\begin{figure}
\centering
\includegraphics[keepaspectratio=true,width=0.5\textwidth]{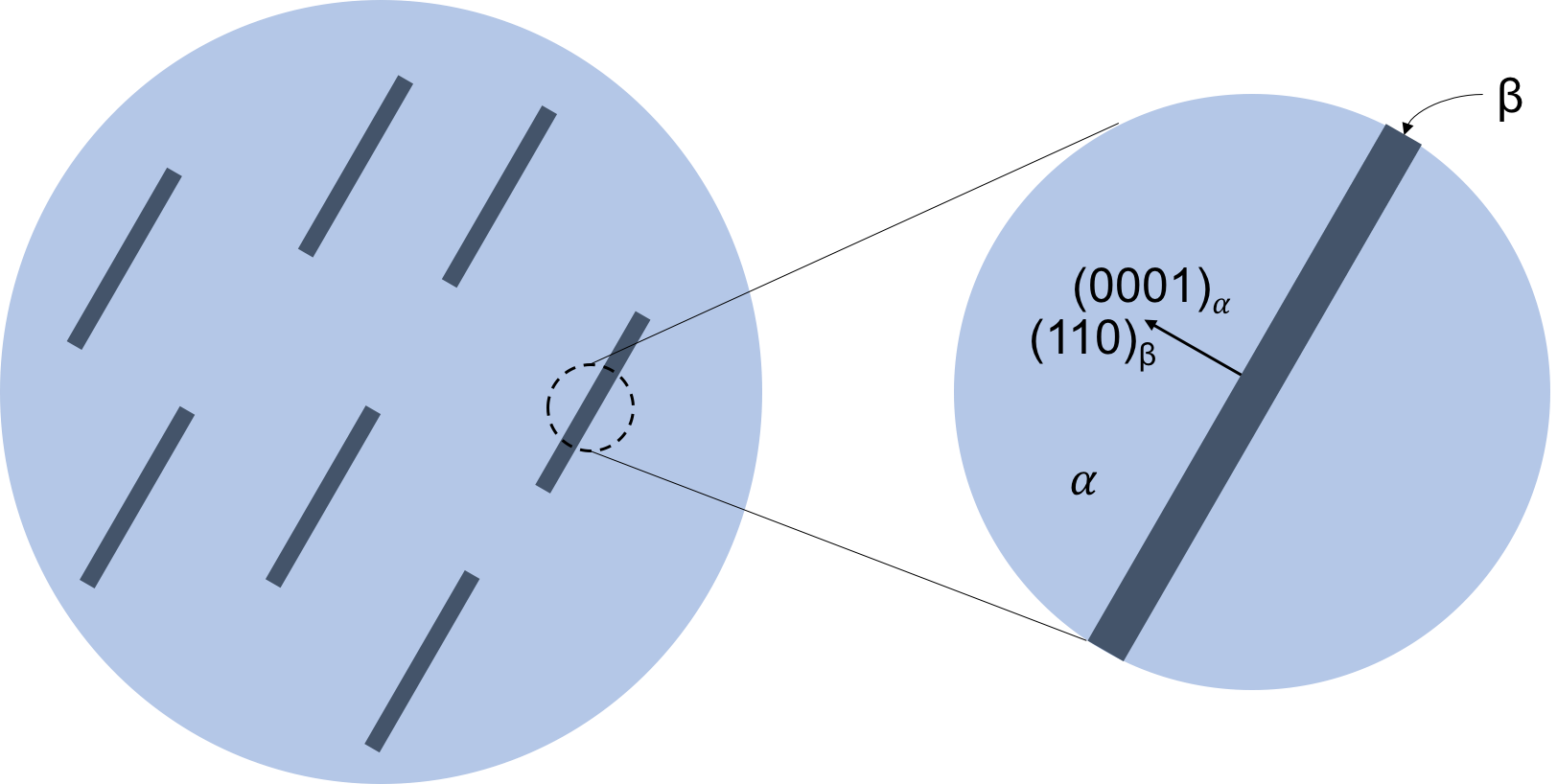}
\caption{Schematic of the microstructure of Mg-Al alloy under study. The plate like $\beta$ phase precipitates are dispersed in $\alpha$ phase. The normal to the interface is in the $(0001)$ direction ($c$ axis) in the $\alpha$ phase and $(110)$ direction in the $\beta$ phase.}\label{Fig:Microstructure}
\end{figure}
We analyze an alloy whose microstructure is shown in Fig. \ref{Fig:Microstructure}, where the plate like $\beta$ phase precipitates are dispersed in an $\alpha$ phase matrix. Assuming the volume fraction of the $\beta$ phase precipitates is much smaller than the volume fraction of the $\alpha$ phase, the strain field in the $\alpha$ phase ($\epsilon^{\alpha}$) is constant and same as the applied strain. At the interface of the $\alpha$ and $\beta$ phase, the following Hadamard jump conditions on stress ($\left[| {\bsigma}^{\alpha,\beta}|\right].\hat{{\bf n}}=0$) and strains ($\left[|{\be}^{\alpha,\beta}|\right].\hat{{\bf t}}=0$) hold. With reference to the $\alpha$ phase, the normal direction $\hat{{\bf n}}$ coincides with the $c$ axis, and the tangential directions $\hat{{\bf t}}$ are along $a$ and $b$ axis. As the strain in the $\alpha$ phase is known, we use the jump conditions, the expression of transformation strain in (\ref{Eqn:TS}), and the orientation relationship to solve for the strains in the $\beta$ phase ($\be^{\beta}$).
   
The grand potential $\Phi^{\alpha,\beta}$ is defined as
\begin{equation}
\Phi^{\alpha,\beta} = \mathcal{F}^{\alpha,\beta} - \mu c^{\alpha,\beta} - \langle\bsigma\rangle\be^{\alpha,\beta} \,\,,
\end{equation}
where the chemical potential $\mu$ is defined as
\begin{eqnarray}\label{Eqn:ChemPot}
\mu[\be^{\beta}, \be^{\alpha}, T]  =  \frac{\partial \mathcal{F}^{\beta} \left[ c=c^{\beta};\be^{\beta},T \right]}{\partial c}  =  \frac{\partial \mathcal{F}^{\alpha} \left[ c=c^{\alpha};\be^{\alpha},T \right]}{\partial c} \,\,.
\end{eqnarray}
The difference in grand potential is $\left[|\Phi^{\alpha,\beta}|\right] = \Phi^{\beta}-\Phi^{\alpha}$. When $\left[|\Phi^{\alpha,\beta}|\right] = 0$, the system is in equilibrium, and the precipitate will neither grow nor shrink. When $\left[|\Phi^{\alpha,\beta}|\right] < 0$, the precipitate will grow, and when $\left[|\Phi^{\alpha,\beta}|\right] > 0$, the precipitate will shrink.

In our calculations, we consider an alloy with 95 at. percent Mg and 5 at. percent Al. Figs. \ref{Fig:GPvol}-\ref{Fig:GPb} show the contour plots of the difference in grand potential as a function of temperature and applied strain. From these plots, we observe that even at zero applied strain, $\Phi^{\beta}<\Phi^{\alpha}$ at all temperatures between 0 to 500 K, thereby implying spontaneous precipitation. This is because the concentration of Al present in this alloy is beyond the equilibrium solubility limits of Al in Mg. Next to determine the role of applied strain, we consider four cases, where compressive and tensile strains are applied volumetric, and along the $c$, $a$ and $b$ axes, respectively. From Fig. \ref{Fig:GPvol}, we observe that compressive volumetric strains lower $\left[|\Phi^{\alpha,\beta}|\right]$, whereas tensile volumetric strains increase $\left[|\Phi^{\alpha,\beta}|\right]$. From Fig.\ref{Fig:GPc}, we observe that compressive strains along the $c$ axis further lower $\left[|\Phi^{\alpha,\beta}|\right]$, increasing the energetic favourability of $\beta$ phase over $\alpha$ phase, thereby promoting the growth of precipitates. Tensile strains along the $c$ axis direction decrease the energetic favourability of $\beta$ phase over $\alpha$ phase, thereby impeding the growth of precipitates. Surprisingly, Figs. \ref{Fig:GPa} and \ref{Fig:GPb} show us that applied strains along $a$ and $b$ axes have negligible influence on the difference in the grand potential, and hence will not influence the growth or shrinkage of the precipitates.          

Figs. \ref{Fig:GPc}-\ref{Fig:GPb} also show the influence of temperature on the difference in grand potential. From a thermodynamic point of view, increase in temperature decreases the energetic favourability of $\beta$ phase over $\alpha$ phase, thereby impeding the growth of precipitates. This can also be understood by looking at Fig. \ref{Fig:Schematic}, where the solubility of Al in Mg increases in temperature.

\begin{figure}
\centering
\subfloat[volumetric]{\includegraphics[keepaspectratio=true,width=0.4\textwidth]{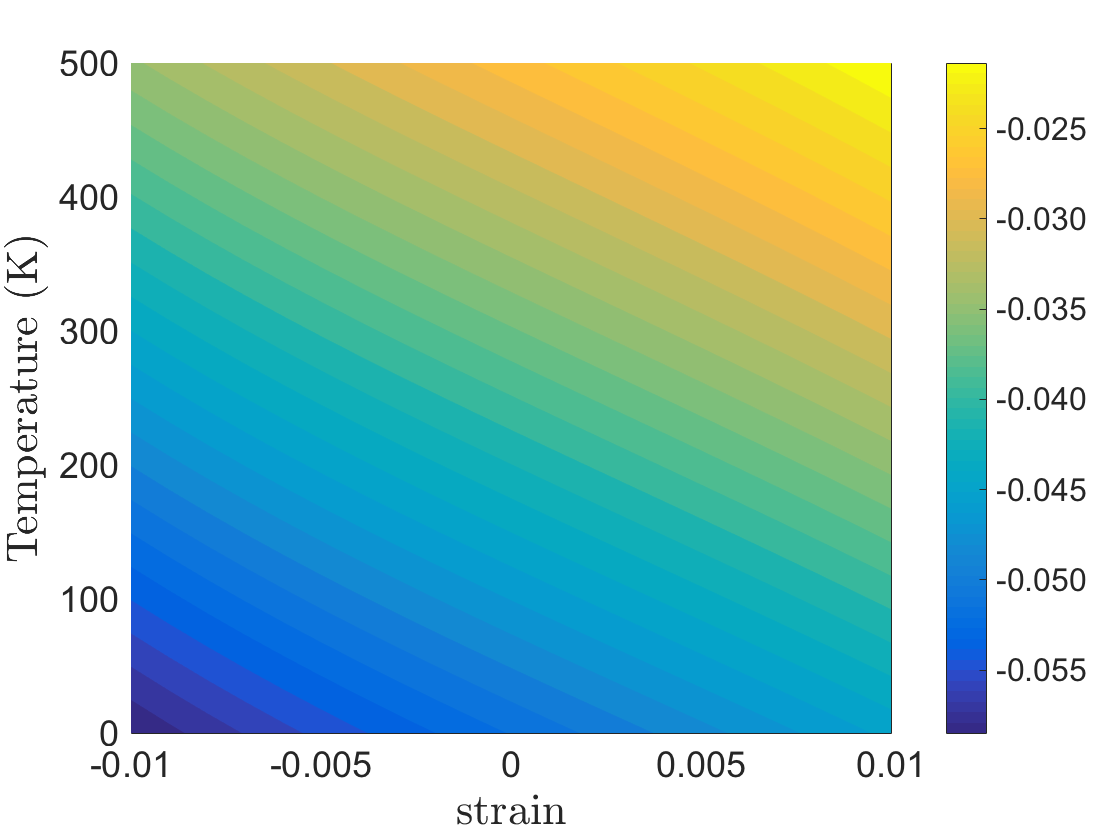}\label{Fig:GPvol}} 
\subfloat[c-axis]{\includegraphics[keepaspectratio=true,width=0.4\textwidth]{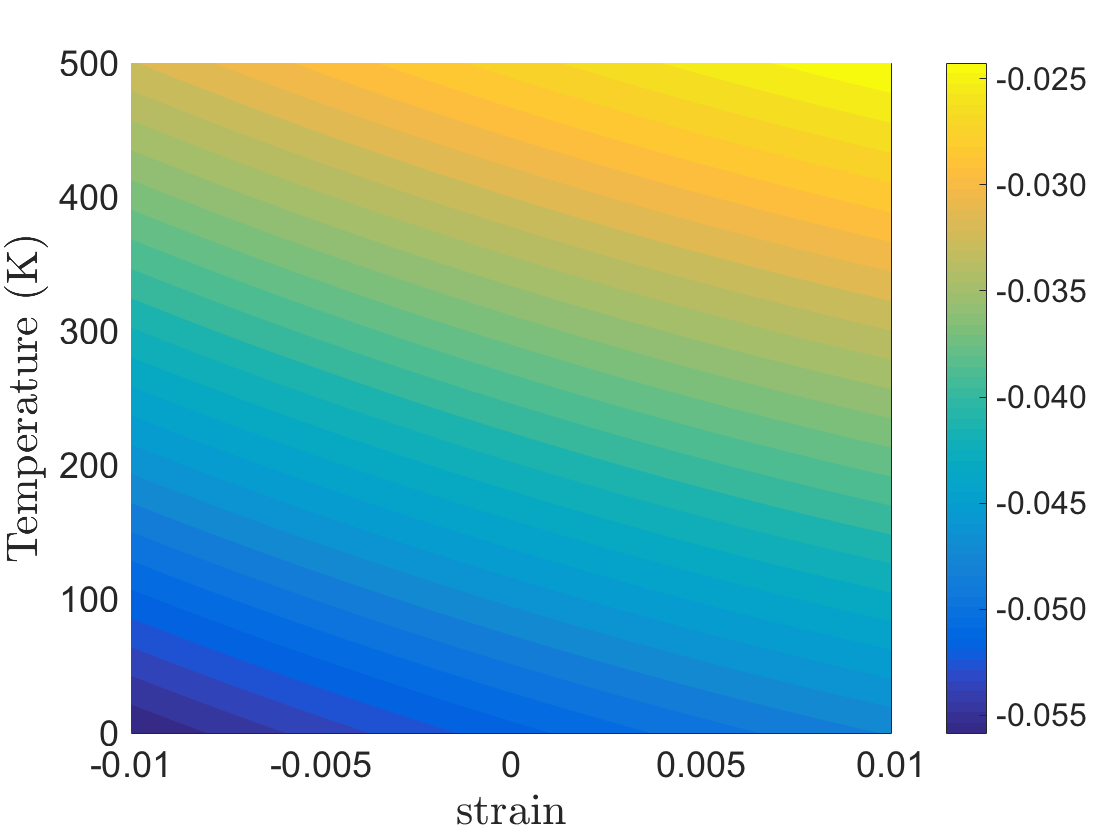}\label{Fig:GPc}} \\
\subfloat[a-axis]{\includegraphics[keepaspectratio=true,width=0.4\textwidth]{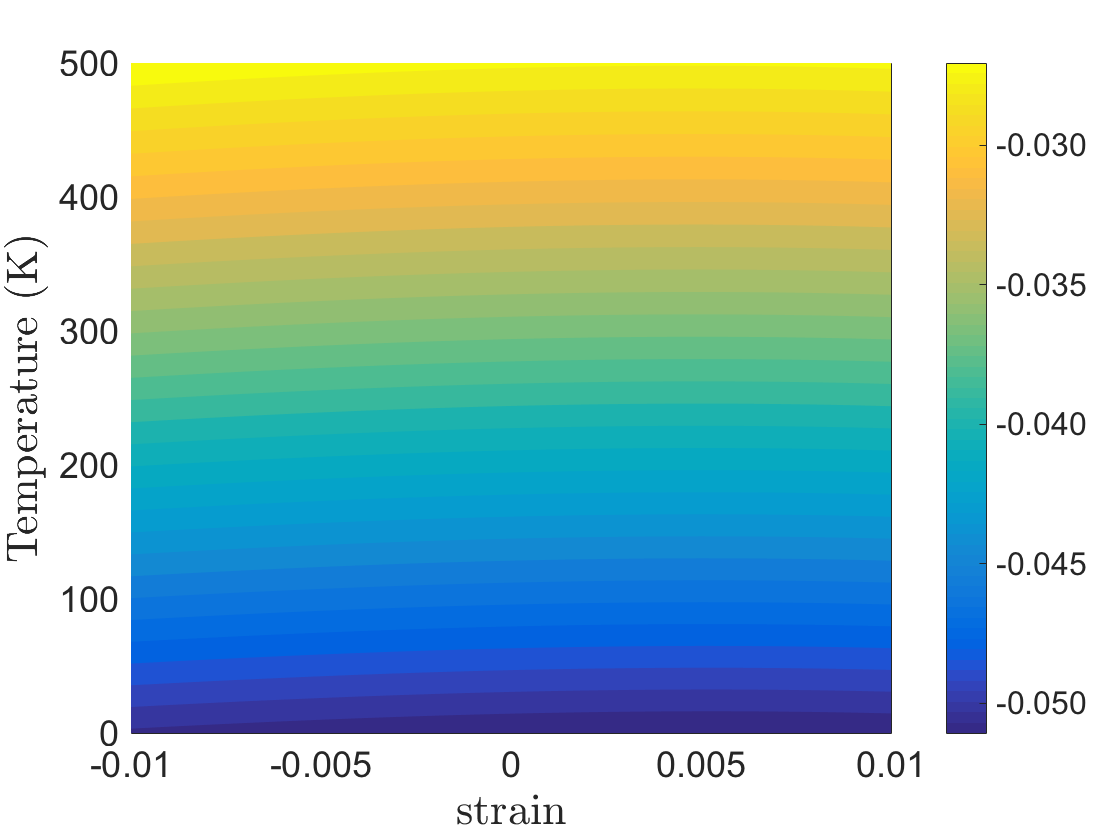}\label{Fig:GPa}} 
\subfloat[b-axis]{\includegraphics[keepaspectratio=true,width=0.4\textwidth]{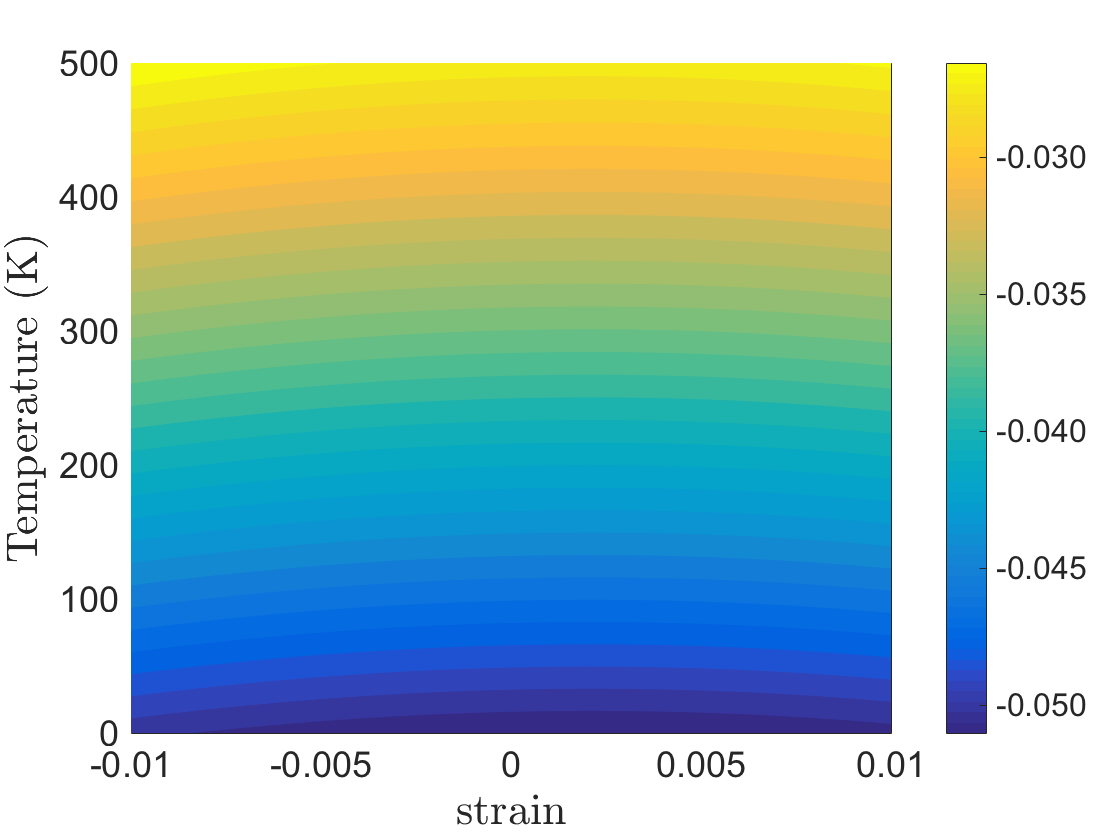}\label{Fig:GPb}} 
\caption{Contours of the values of the difference in grand potential ($\left[|\Phi^{\alpha,\beta}|\right]$) across the $\beta$ and $\alpha$ phases as a function of applied volumetric strain and strain in $c$, $a$ and $b$ axes in the $\alpha$ phase, and temperature.}
\end{figure}

\paragraph*{Equilibrium chemical potential and solubility limits} 
We calculate the influence of applied strain and temperature on the solubility limits of Al in Mg ($c_{\alpha}^0$), Mg in Mg$_{17}$Al$_{12}$ ($c_{\beta}^0$), and the chemical potential $\mu$ at equilibrium. At equilibrium, the difference in grand potential is zero ($\left[|\Phi^{\alpha,\beta}|\right] = 0$), and the grand potential in the individual phases is minimum with respect to the concentration. These are computed from the slope of the common tangent between the $\beta$ phase solutions and $\alpha$ phase solutions. This is given by

\begin{eqnarray}\label{Eqn:CTConditions}
\mu[\be^{\beta},\be^{\alpha}, T] & = & \frac{\partial \mathcal{F}^{\beta} \left[ c=c_{\beta}^0;\be^{\beta},T \right]}{\partial c} \nonumber \\ & = & \frac{\partial \mathcal{F}^{\alpha} \left[ c=c_{\alpha}^0;\be^{\alpha},T \right]}{\partial c} \nonumber \\ &=& \frac{\left[| \mathcal{F}^{\alpha,\beta} - \langle\bsigma\rangle\be^{\alpha,\beta}|\right]}{\left[|c_{\alpha,\beta}^0 | \right]} \,\,.
\end{eqnarray}

We numerically solve Eqn. \ref{Eqn:CTConditions} using the Levenberg-Marquardt \cite{Levenberg:1944,Marquardt:1963} algorithm. Figs. \ref{Fig:Calphavol}-\ref{Fig:Calphab}, show the influence of applied volumetric strains, applied uniaxial strains in $a$, $b$ and $c$ directions in the $\alpha$ phase, and temperature on the equilibrium solubility limits of Al in Mg. From these contour plots, we see that compressive volumetric strains and compressive strains along the $c$ axis decrease the solubility of Al in Mg, whereas tensile volumetric strains and strains along the $c$ axis increase the solubility of Al in Mg. Strains along $a$ and $b$ axis directions do not influence the solubility of Al in Mg. Temperature has a strong influence on the solubility limits of Al in Mg. It is known from the phase diagram Fig. \ref{Fig:Schematic}, that in absence of strain, increase in temperature will increase the solubility of Al in Mg. Figs. \ref{Fig:Calphavol}-\ref{Fig:Calphab}, tell us that this holds true for strained cases as well, and also gives us the solubility limits in strain-temperature space.  
 
\begin{figure}
\centering
\subfloat[volumetric]{\includegraphics[keepaspectratio=true,width=0.4\textwidth]{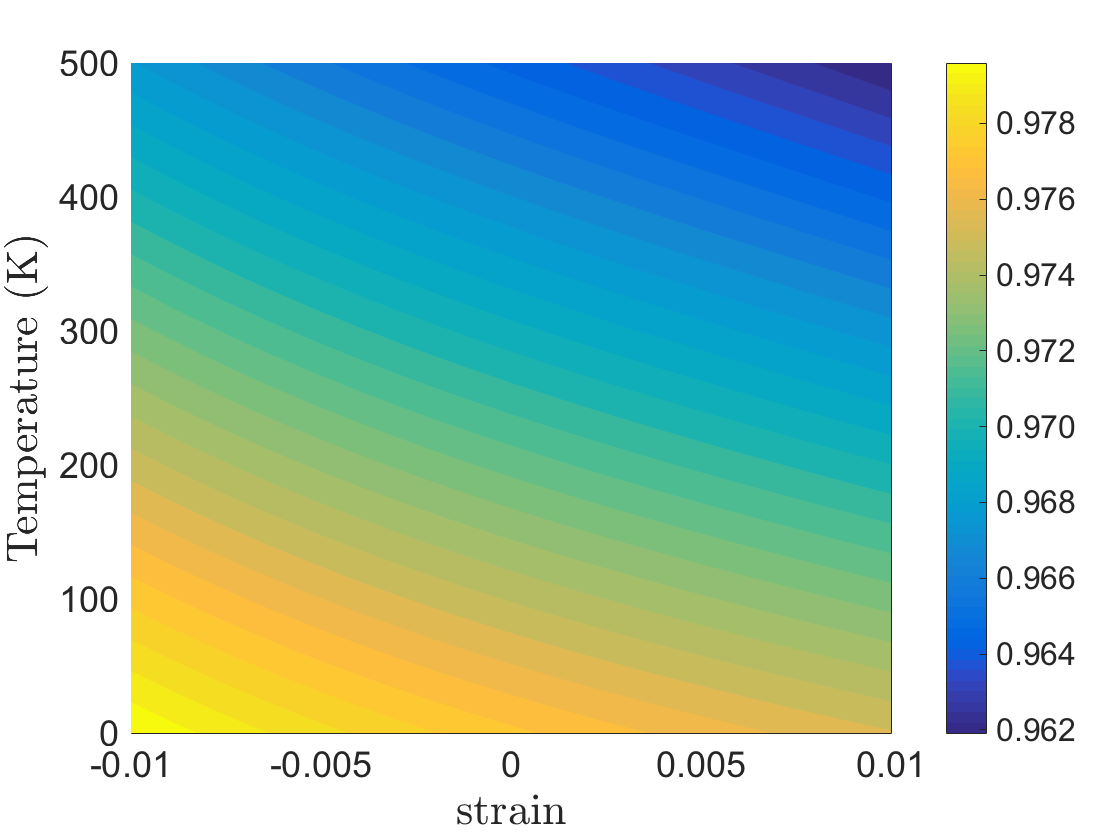}\label{Fig:Calphavol}}
\subfloat[c-axis]{\includegraphics[keepaspectratio=true,width=0.4\textwidth]{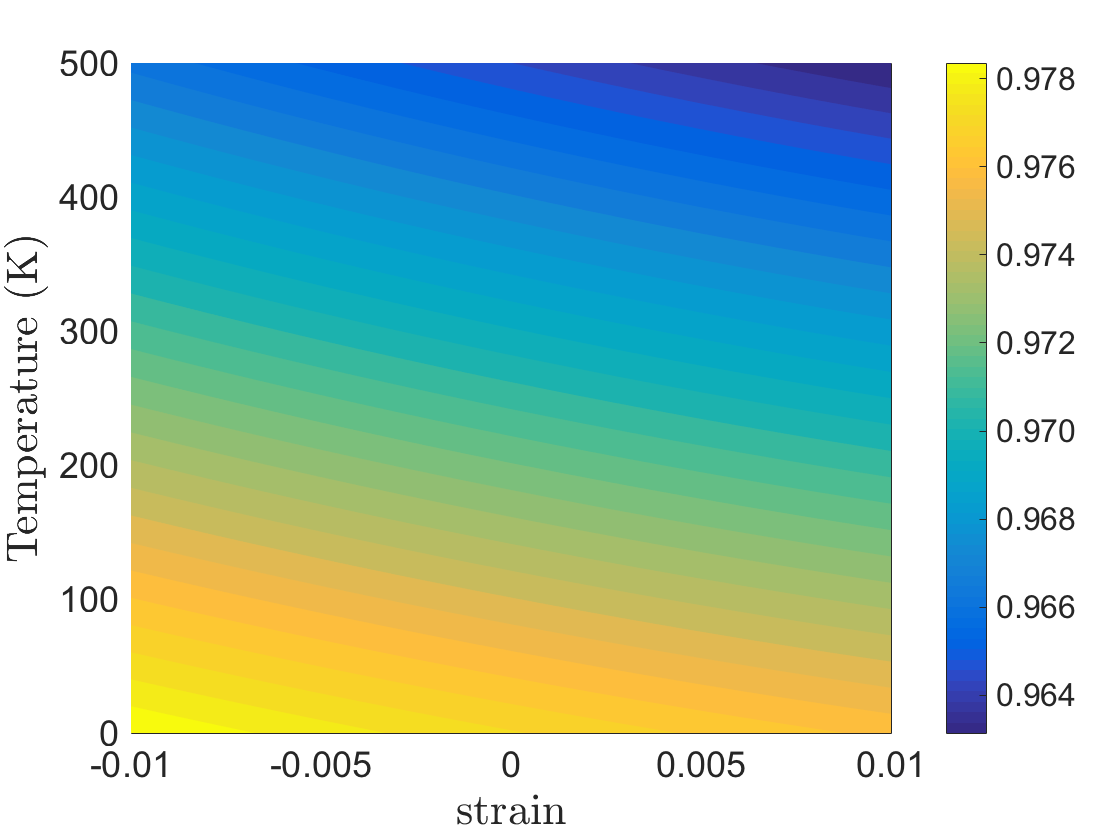}\label{Fig:Calphac}} \\
\subfloat[a-axis]{\includegraphics[keepaspectratio=true,width=0.4\textwidth]{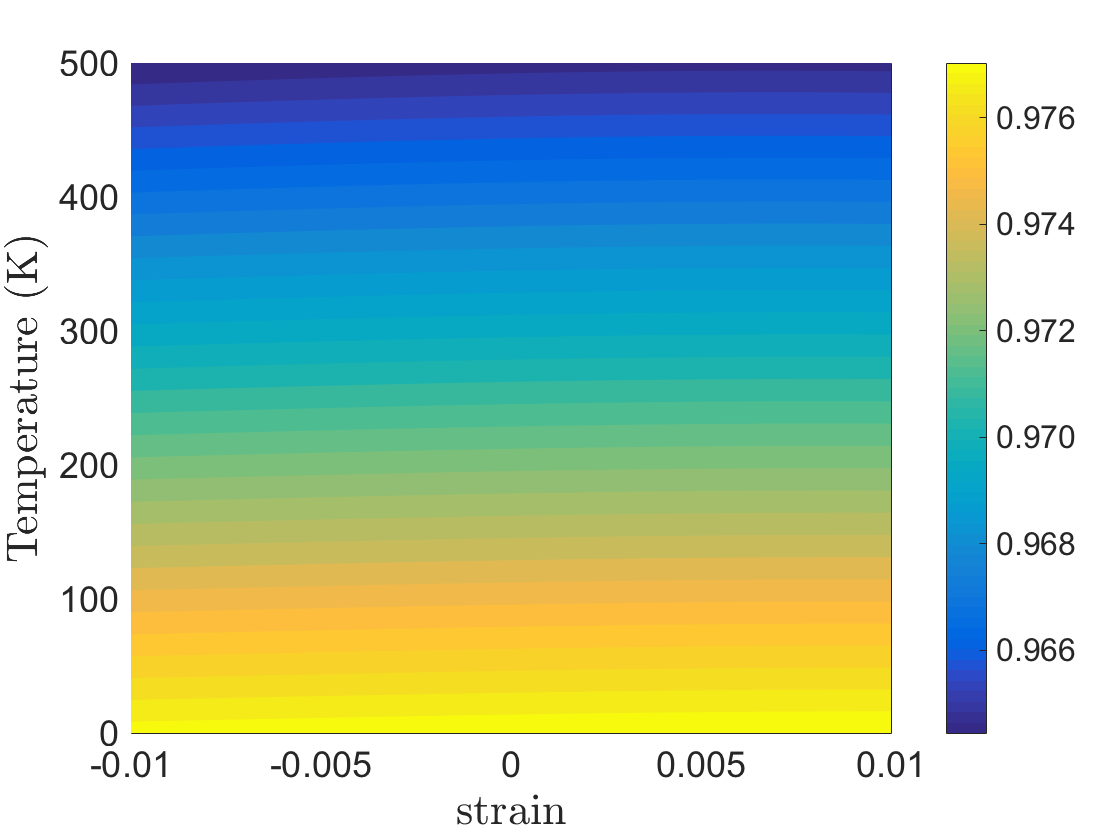}\label{Fig:Calphaa}} 
\subfloat[b-axis]{\includegraphics[keepaspectratio=true,width=0.4\textwidth]{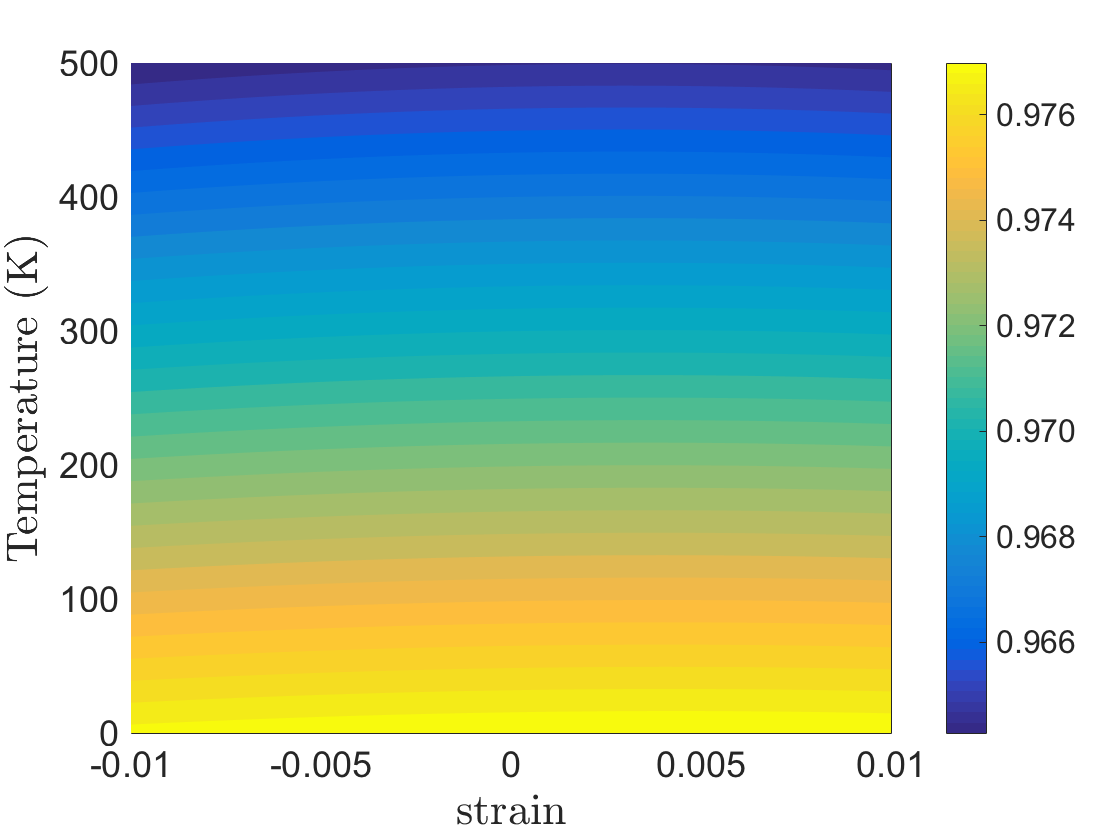}\label{Fig:Calphab}} 
\caption{Contours of the values of the solubility limits ($c_{\alpha}^0$) of Al in Mg as a function of applied volumetric strain and strain in $c$, $a$ and $b$ axes in the $\alpha$ phase, and temperature.}
\end{figure}     
 
Figs. \ref{Fig:Cbetavol}-\ref{Fig:Cbetab}, show the influence of applied volumetric and uniaxial strains in $a$, $b$ and $c$ directions in the $\alpha$ phase, and temperature on the equilibrium solubility limits ($c_{\beta}^0$) of Mg in Mg$_{17}$Al$_{12}$. These plots tell us that there is slight variation in $c_{\beta}^0$ with strain and temperature. Compressive strains along the $c$ axis increase the solubility of Mg in Mg$_{17}$Al$_{12}$, whereas tensile strains in the $c$ axis direction decrease it. For the volumetric strains, and strains applied along the $a$ and $b$ axis directions, both tensile and compressive strains decrease the solubility. Increase in temperature also increase the solubility. Overall, the variation in solubility of Mg in Mg$_{17}$Al$_{12}$ with applied strain and temperature is small when compared to the variation in solubility of Al in Mg.   

\begin{figure}
\centering
\subfloat[volumetric]{\includegraphics[keepaspectratio=true,width=0.4\textwidth]{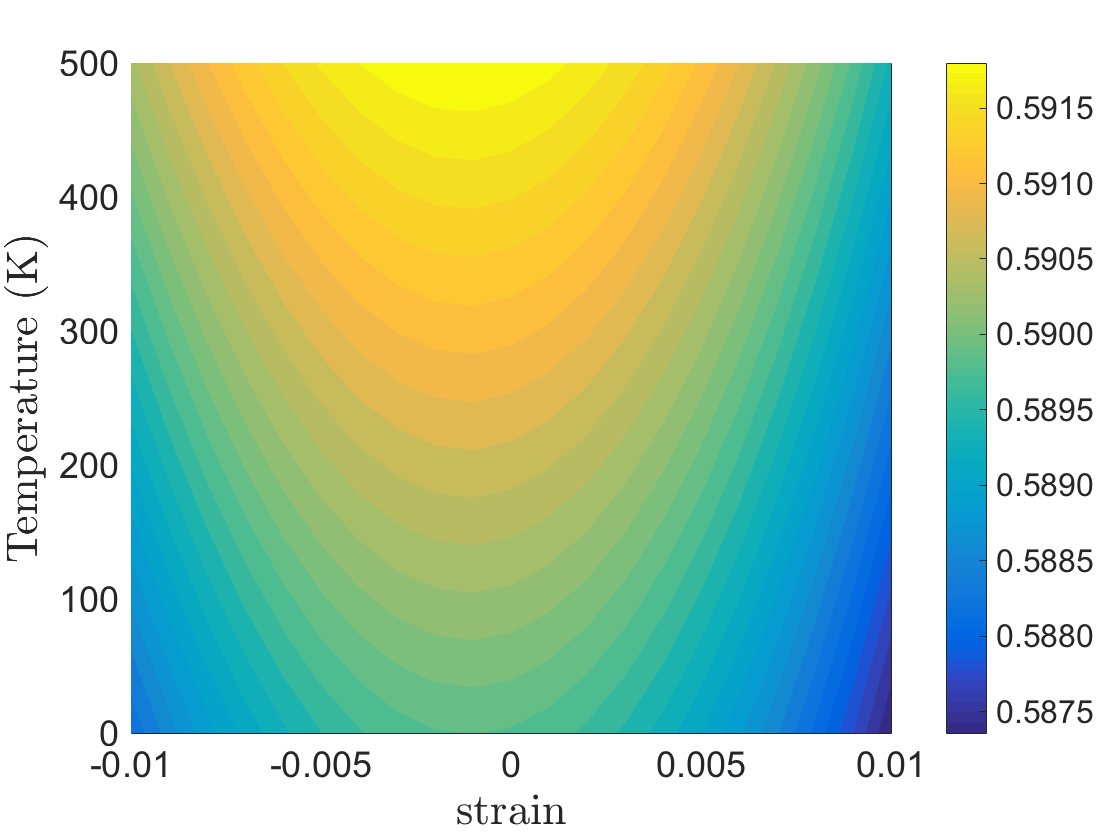}\label{Fig:Cbetavol}}
\subfloat[c-axis]{\includegraphics[keepaspectratio=true,width=0.4\textwidth]{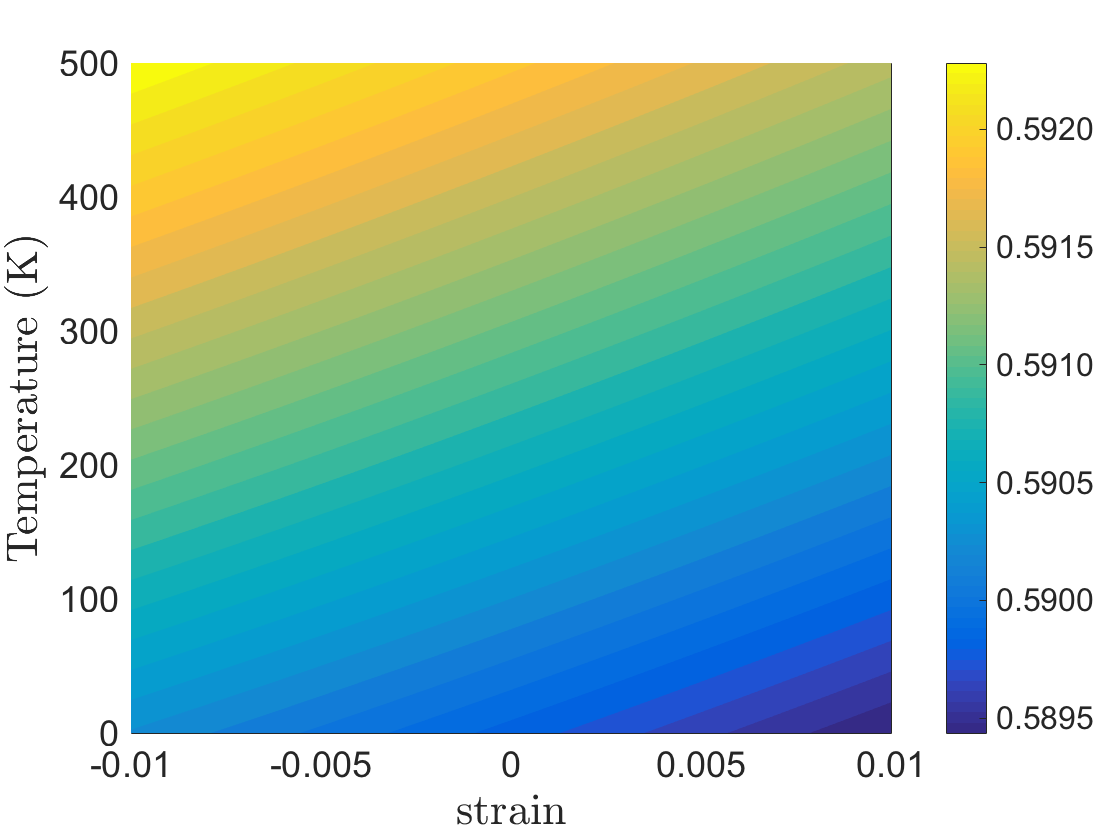}\label{Fig:Cbetac}} \\
\subfloat[a-axis]{\includegraphics[keepaspectratio=true,width=0.4\textwidth]{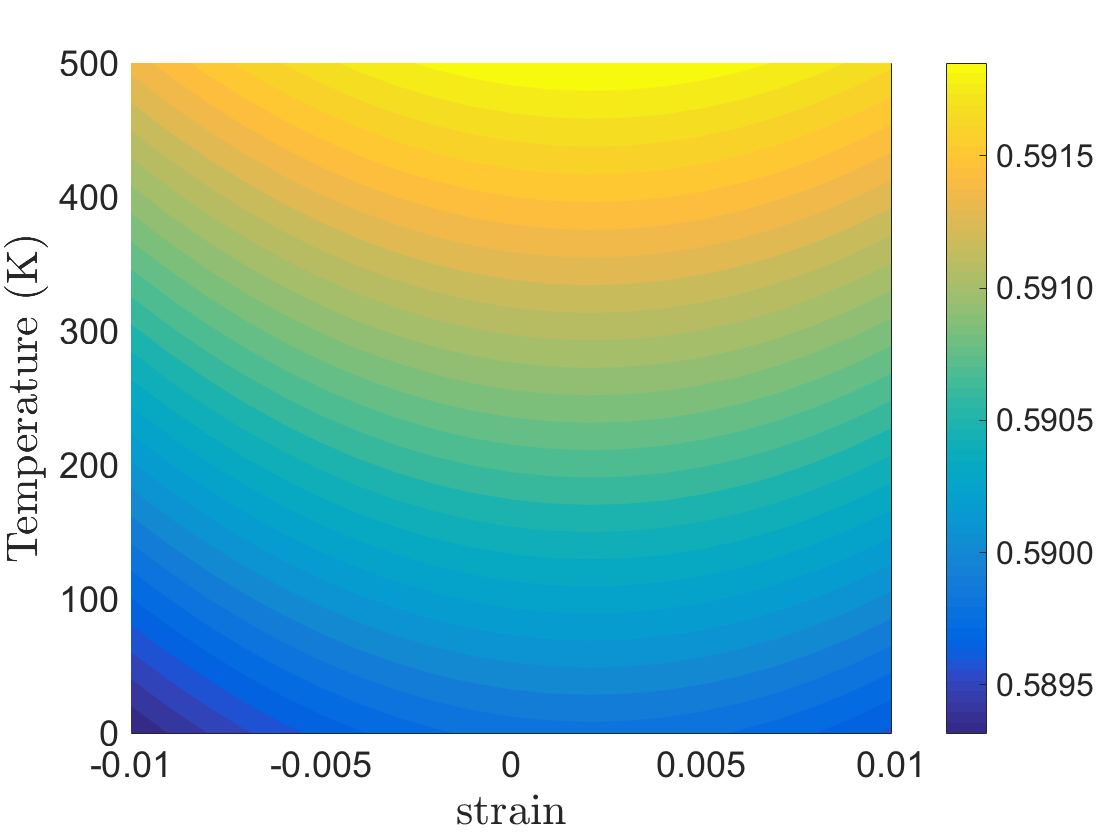}\label{Fig:Cbetaa}} 
\subfloat[b-axis]{\includegraphics[keepaspectratio=true,width=0.4\textwidth]{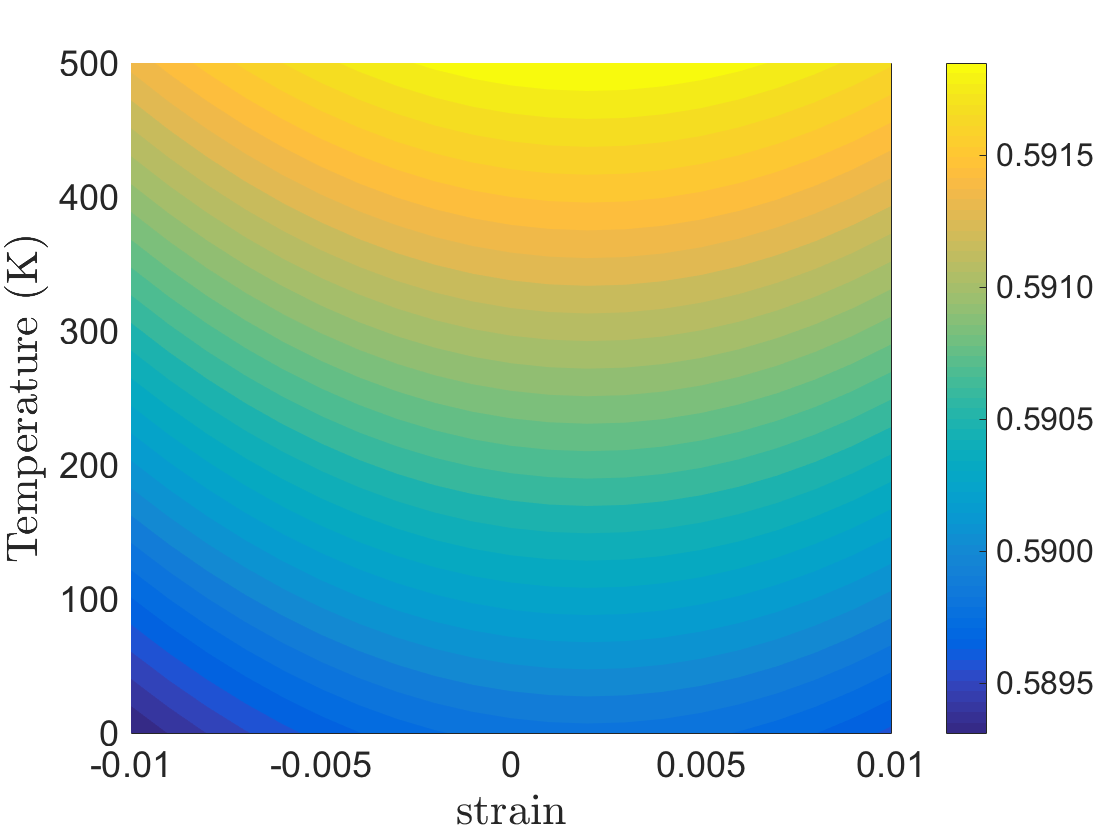}\label{Fig:Cbetab}} 
\caption{Contours of the values of the solubility limits ($c_{\beta}^0$) of Mg in Mg$_{17}$Al$_{12}$  as a function of applied volumetric strain and strain in $c$, $a$ and $b$ axes in the $\alpha$ phase, and temperature.}
\end{figure}      
  
\begin{figure}
\centering
\subfloat[volumetric]{\includegraphics[keepaspectratio=true,width=0.4\textwidth]{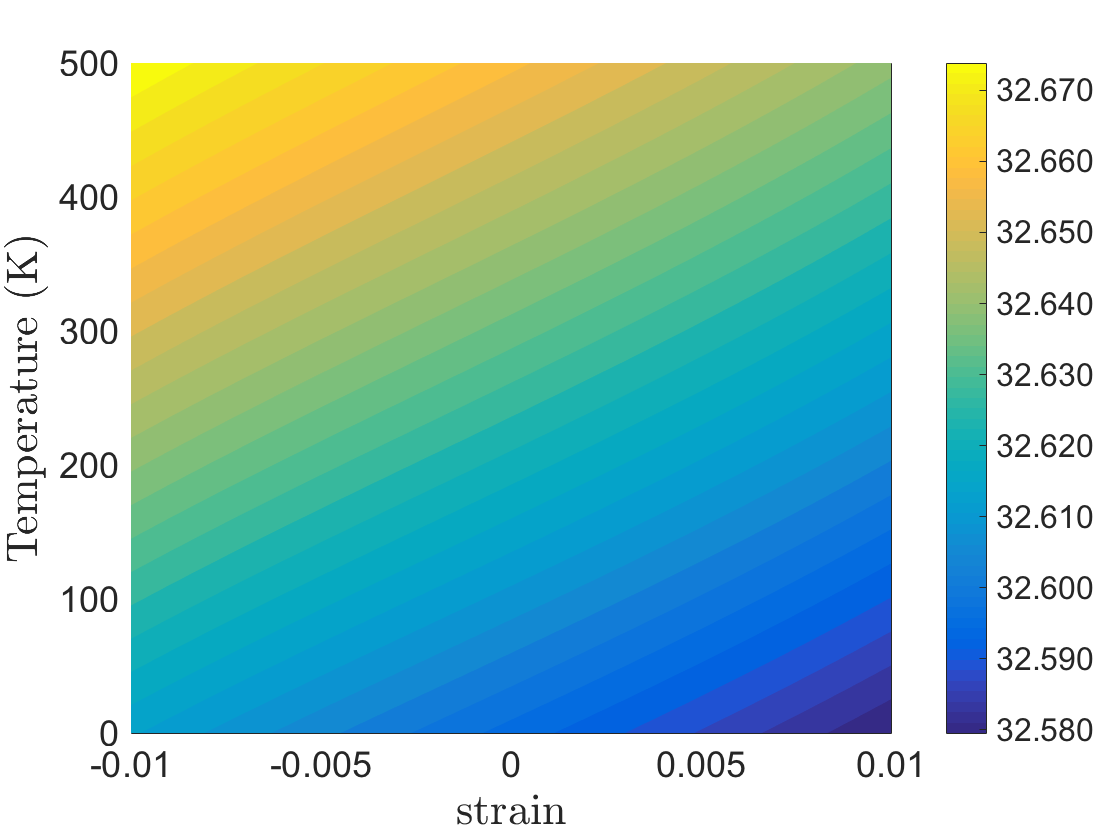}\label{Fig:Potentialvol}}
\subfloat[c-axis]{\includegraphics[keepaspectratio=true,width=0.4\textwidth]{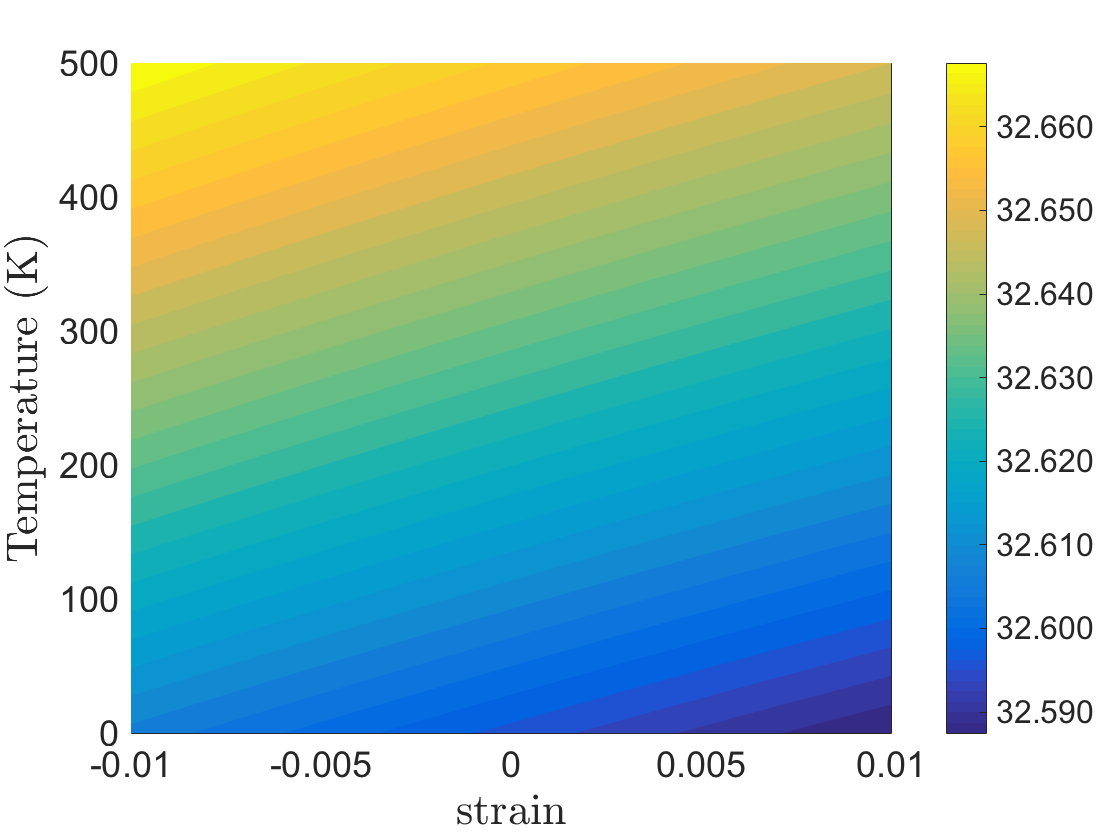}\label{Fig:Potentialc}} \\
\subfloat[a-axis]{\includegraphics[keepaspectratio=true,width=0.4\textwidth]{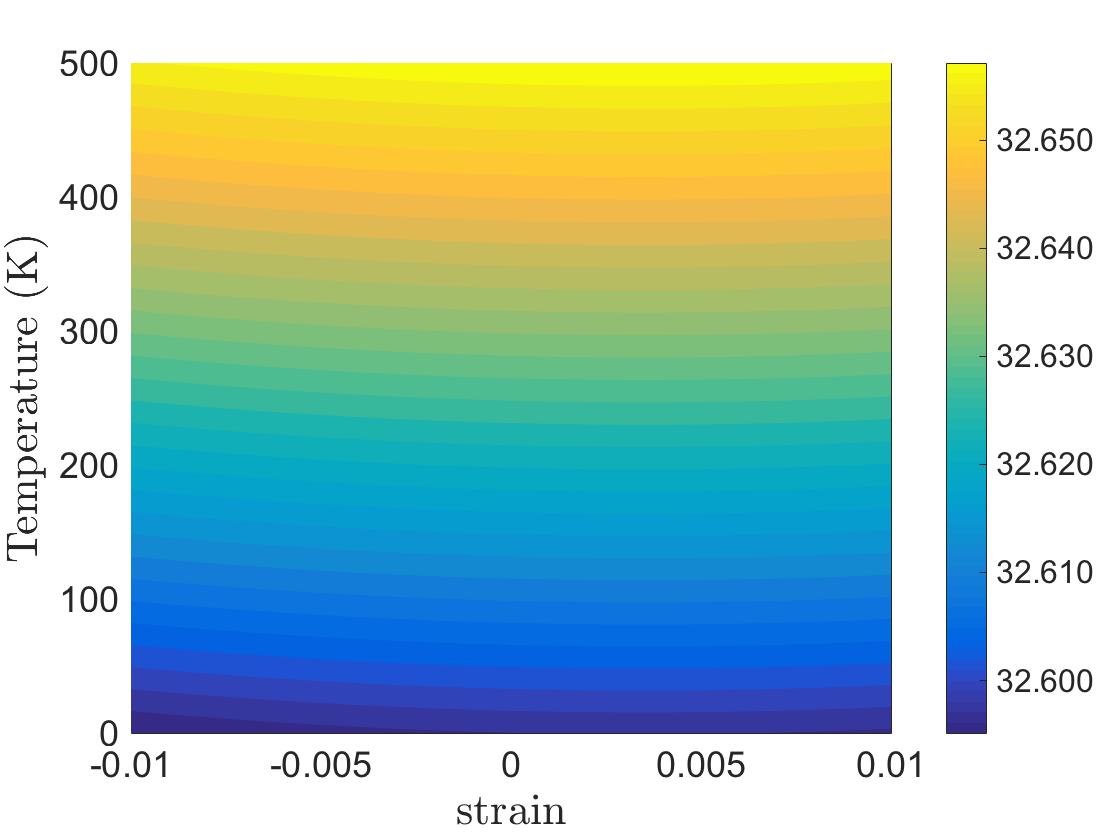}\label{Fig:Potentiala}} 
\subfloat[b-axis]{\includegraphics[keepaspectratio=true,width=0.4\textwidth]{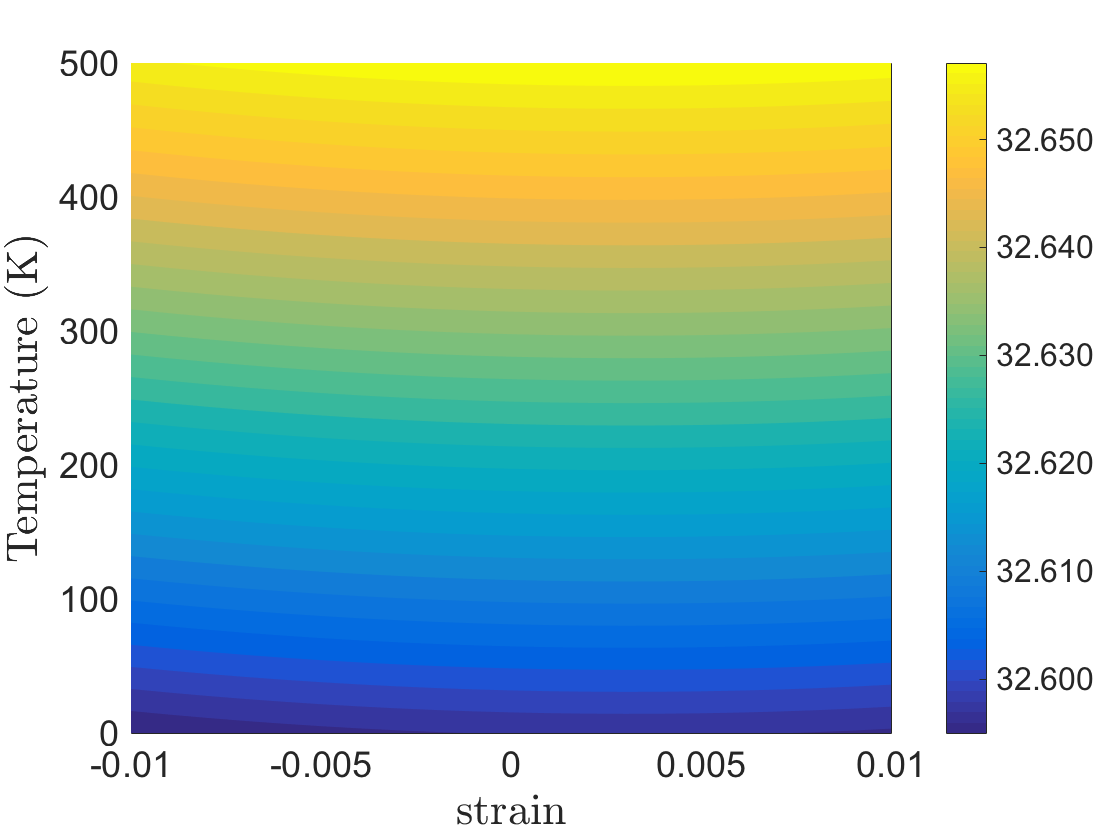}\label{Fig:Potentialb}} 
\caption{Contours of the values of the chemical potential ($\mu$) at equilibrium as a function of applied volumetric strain and strain in $c$, $a$ and $b$ axes in the $\alpha$  phase, and temperature.}
\end{figure}      
In Figs. \ref{Fig:Potentialvol}-\ref{Fig:Potentialb}, we plot the contours of the values of the chemical potential $\mu$ under the influence of temperature and applied volumetric and uniaxial strains in $a$, $b$ and $c$ directions in the $\alpha$ phase, respectively. From these plots, we conclude that the chemical potential at equilibrium is significantly influenced by temperature and the nature of strain, where compressive volumetric strains and strains applied along the $c$ axis direction increases the equilibrium chemical potential whereas tensile strains along these decreases it. Strains applied along the $a$ and $b$ axis directions do not influence the equilibrium chemical potential. Increase in temperature also increases the equilibrium chemical potential. 

Overall, we conclude that temperature and nature of strains have a significant influence on the chemical potential at equilibrium and solubility limits in magnesium-aluminum alloys. While the effect of temperature on these properties at zero strain can be accounted for from the (zero strain) phase diagram (Fig \ref{Fig:Schematic}), the influence of strain on the solubility limits and chemical potential has been unknown prior to this work. As it is common to encounter a combination of thermal and mechanical loads during processing, our data can provide a reliable input to process models to understand precipitation in these alloys.     

\paragraph{Excess free energy}
Finally, we report the influence of thermomechanical loads on the excess free energy when a solid solution with concentration $c$ undergoes phase separation. The change in free energy is taken with respect to pure magnesium and the pure precipitate phase (Mg$_{17}$Al$_{12}$), and is calculated from the free energies of the solid solution, and the individual free energies of magnesium and the precipitate. The individual strains in the solid solution, pure magnesium and the precipitate are $\be$, $\be^{\alpha}$ and $\be^{\beta}$, respectively. At a given temperature $T$, this change in free energy ($\Delta \mathcal{F}^{\alpha,\beta}$) is expressed as
\begin{equation}\label{Eqn:FreeEnergyChange}
\Delta \mathcal{F}^{\alpha,\beta}[c,\be,\be^{\alpha},\be^{\beta},T] = \mathcal{F}^{\alpha,\beta}[c,\be,T]- \lambda(c) \,\mathcal{F}^{\alpha}[c^{\alpha},\be^{\alpha},T]-(1-\lambda(c)) \mathcal{F}^{\beta}[c^{\beta},\be^{\beta},T] \,\,,
\end{equation}
where, $c^{\beta} = 0.586$, $c^{\alpha} = 1$, and the function $\lambda(c)$ = $(c-c^{\beta})/(1-c^{\beta})$ is such that $\lambda(c^{\beta})$=0 and $\lambda(c^{\alpha})$ = 1. The resulting condition on the free energy change is $\Delta \mathcal{F}^{\beta}[c^{\beta};\be^{\beta},T]$ = $\Delta \mathcal{F}^{\alpha}[c^{\alpha},\be^{\alpha},T]$ = $0$, for all temperature. The change in free energy is negative if the free energy of the solid solution is less than the free energy of the phase separated mixture of a precipitate phase and a pure magnesium phase.

On substituting Eqn. \ref{Eqn:EnergyFit} in Eqn. \ref{Eqn:FreeEnergyChange}, we can decompose the free energy change as
\begin{equation}\label{Eqn:FEChange}
\Delta \mathcal{F}^{\alpha,\beta}[c,\be,\be^{\alpha},\be^{\beta},T] = \Delta \mathcal{F}_0^{\alpha,\beta}[c,T] + \Delta \mathcal{F}_{strain}^{\alpha,\beta}[c,\be,\be^{\alpha},\be^{\beta}] \,\,,
\end{equation}
where these terms are given by 
\begin{equation} \label{Eqn:FEChange1}
\Delta \mathcal{F}_0[c,T] = \mathcal{E}_0^{\alpha,\beta}[c] -T S[c] - \lambda(c) (\mathcal{E}_0^{\alpha}[c^{\alpha}]- T S[c^{\alpha}])-(1-\lambda(c)) \left( \mathcal{E}_0^{\beta}[c^{\beta}] - T S[c^{\beta}] \right) \,\,, 
\end{equation}
and
\begin{equation}\label{Eqn:FEChange2}
\Delta \mathcal{F}_{strain}^{\alpha,\beta}[c,\be,\be^{\alpha},\be^{\beta}] = \mathcal{E}_{strain}^{\alpha,\beta}[c,\be] -  \lambda(c) \mathcal{E}_{strain}^{\alpha}[c^{\alpha},\be^{\alpha}] - (1-\lambda(c))\mathcal{E}_{strain}^{\beta}[c^{\beta},\be^{\beta}] \,\,.
\end{equation}
 Notice that the configurational entropy for pure magnesium is zero and therefore entropy at $c$ = 1, is only due to the electronic contribution, i.e. $S(c=1)$ = $S_e(1)$. In Eqn. \ref{Eqn:FEChange1}, the energy $\mathcal{E}_0^{\alpha,\beta}[c]$ is given by Eqn. \ref{Eqn:EnergyFit}, and $\mathcal{E}_{strain}^{\alpha,\beta}[c]$ is the contribution of the strain to the total energy in Eqn. \ref{Eqn:EnergyFit}, i.e. $\mathcal{E}_{strain}^{\alpha,\beta}[c,\epsilon]$ = $\mathcal{E}^{\alpha,\beta}[c,\epsilon] - \mathcal{E}_0^{\alpha,\beta}[c]$.  

\begin{figure}
\subfloat[zero strain]
{\includegraphics[keepaspectratio=true,width=0.35\textwidth]{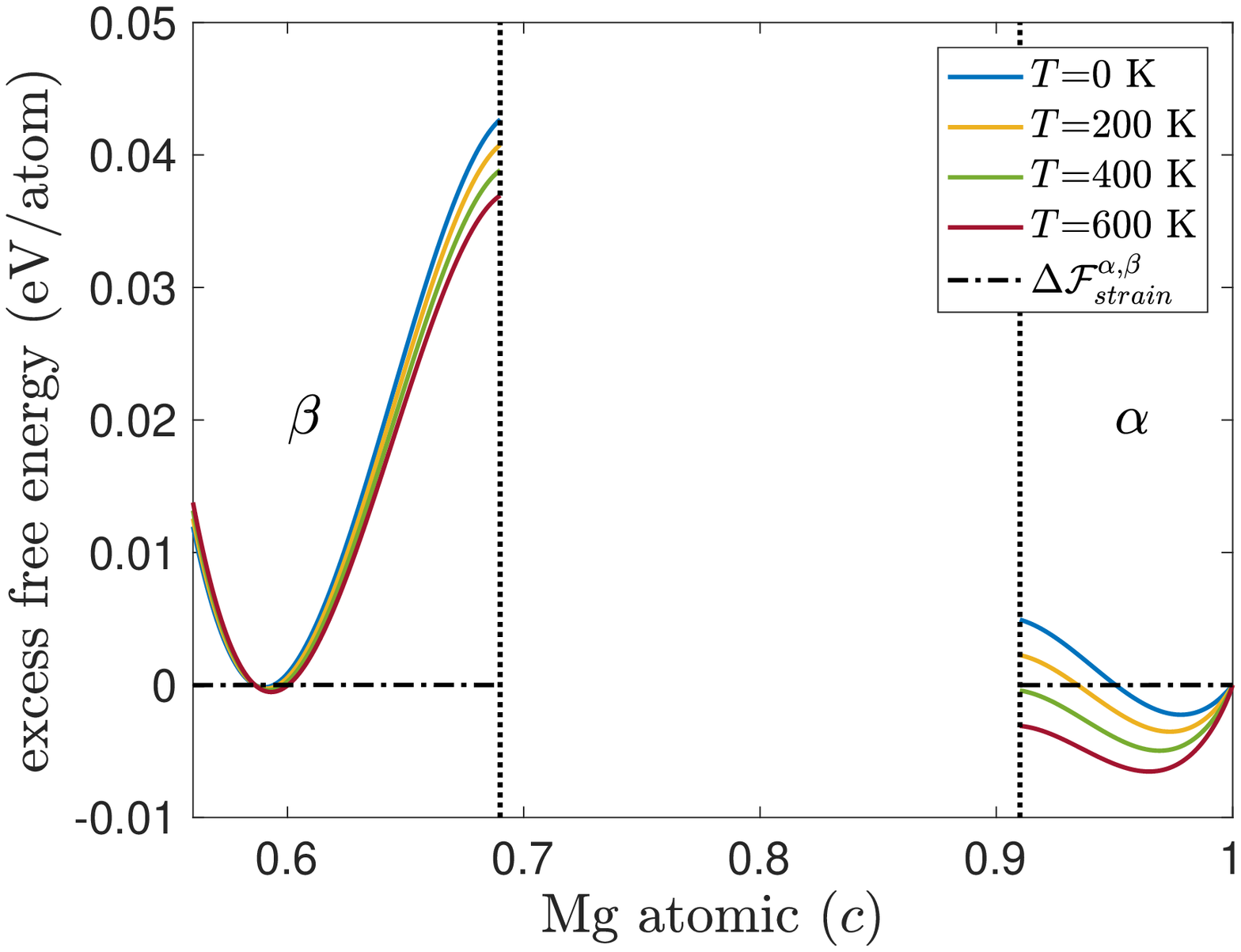}\label{Fig:FEZeroStrain}} 
\subfloat[volumetric compression (-0.01)]
{\includegraphics[keepaspectratio=true,width=0.35\textwidth]{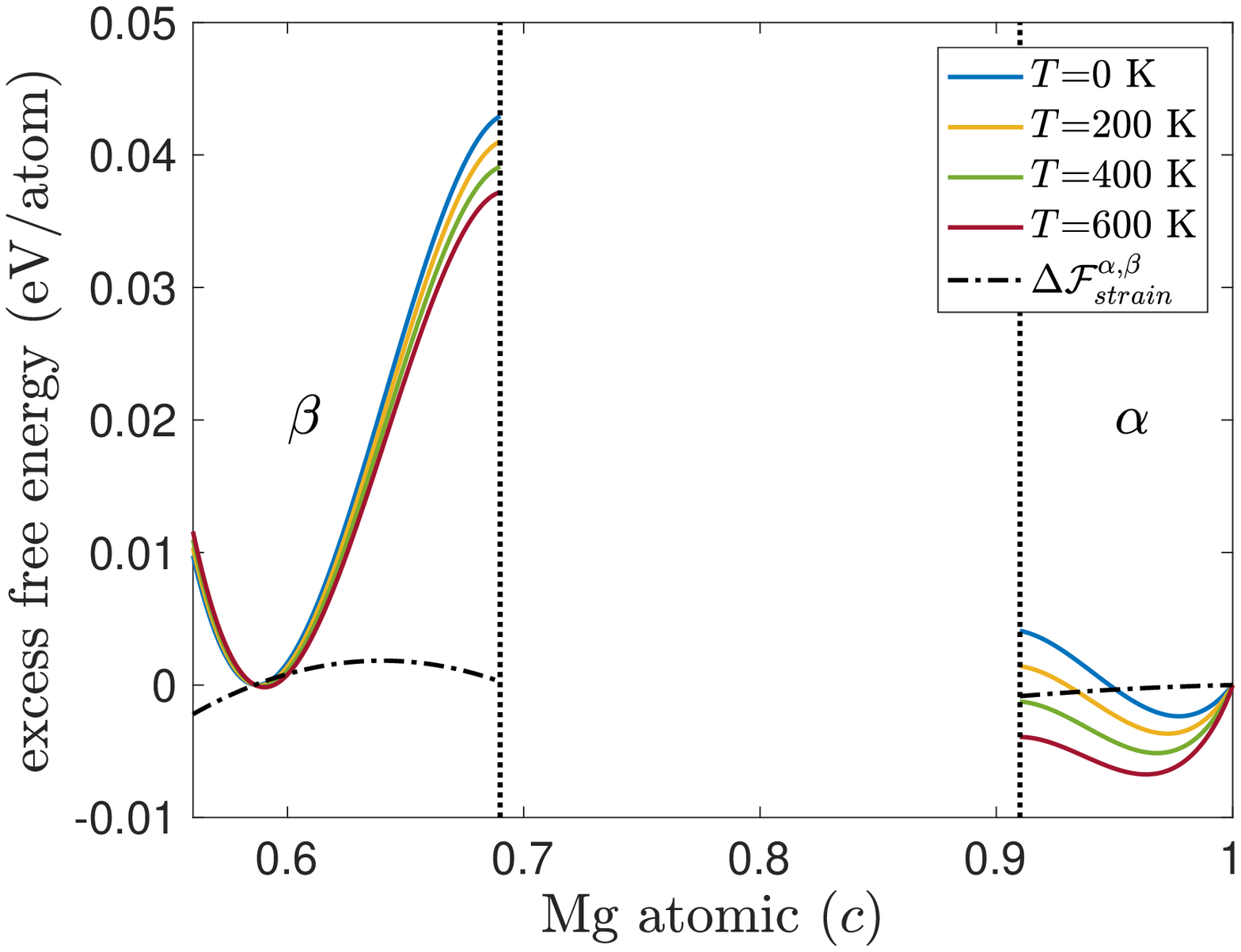}\label{Fig:FEVC}}
\subfloat[uniaxial compression a (-0.03)] 
{\includegraphics[keepaspectratio=true,width=0.35\textwidth]{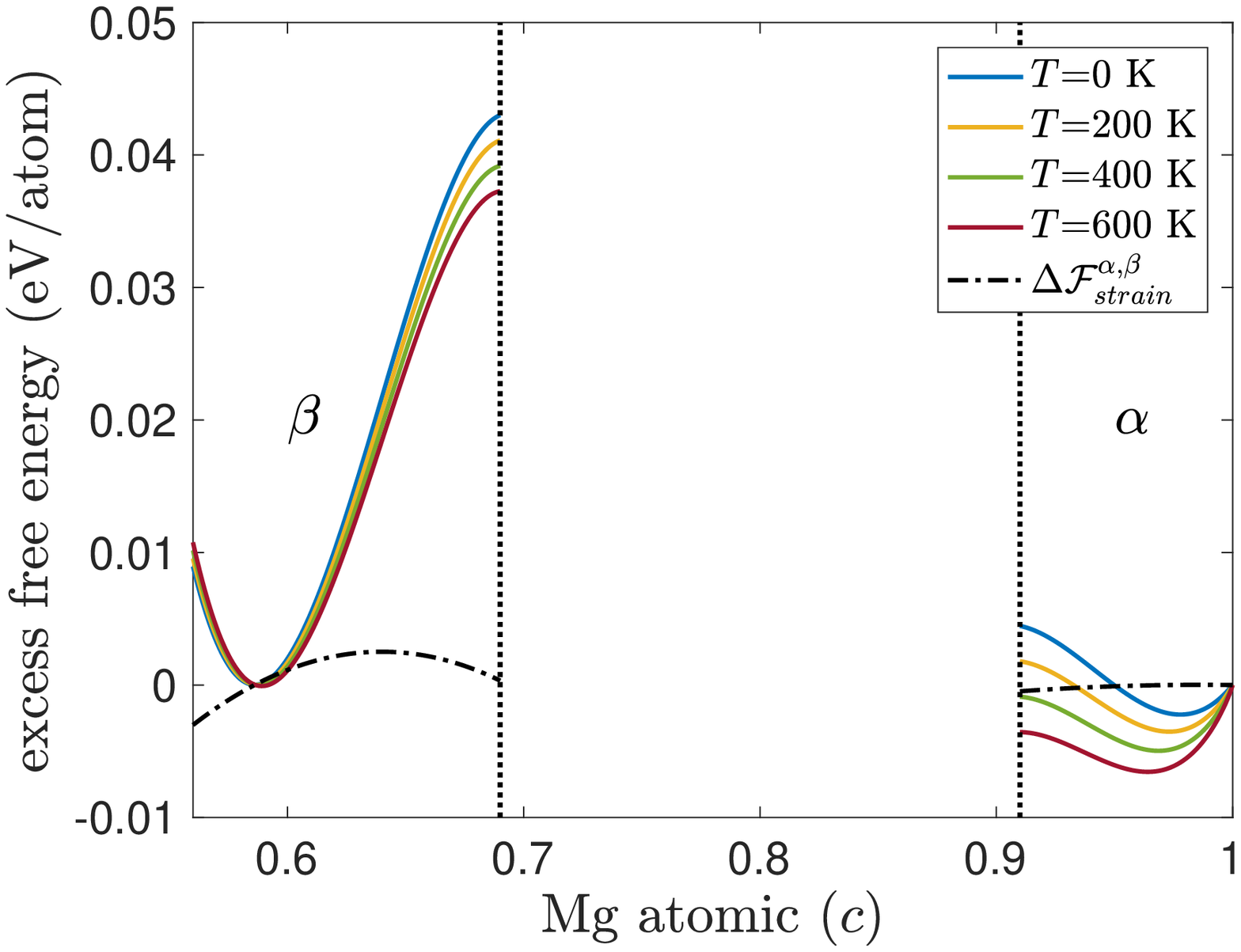}\label{Fig:FEUCa}} \\
\subfloat[uniaxial compression b (-0.03)]
{\includegraphics[keepaspectratio=true,width=0.35\textwidth]{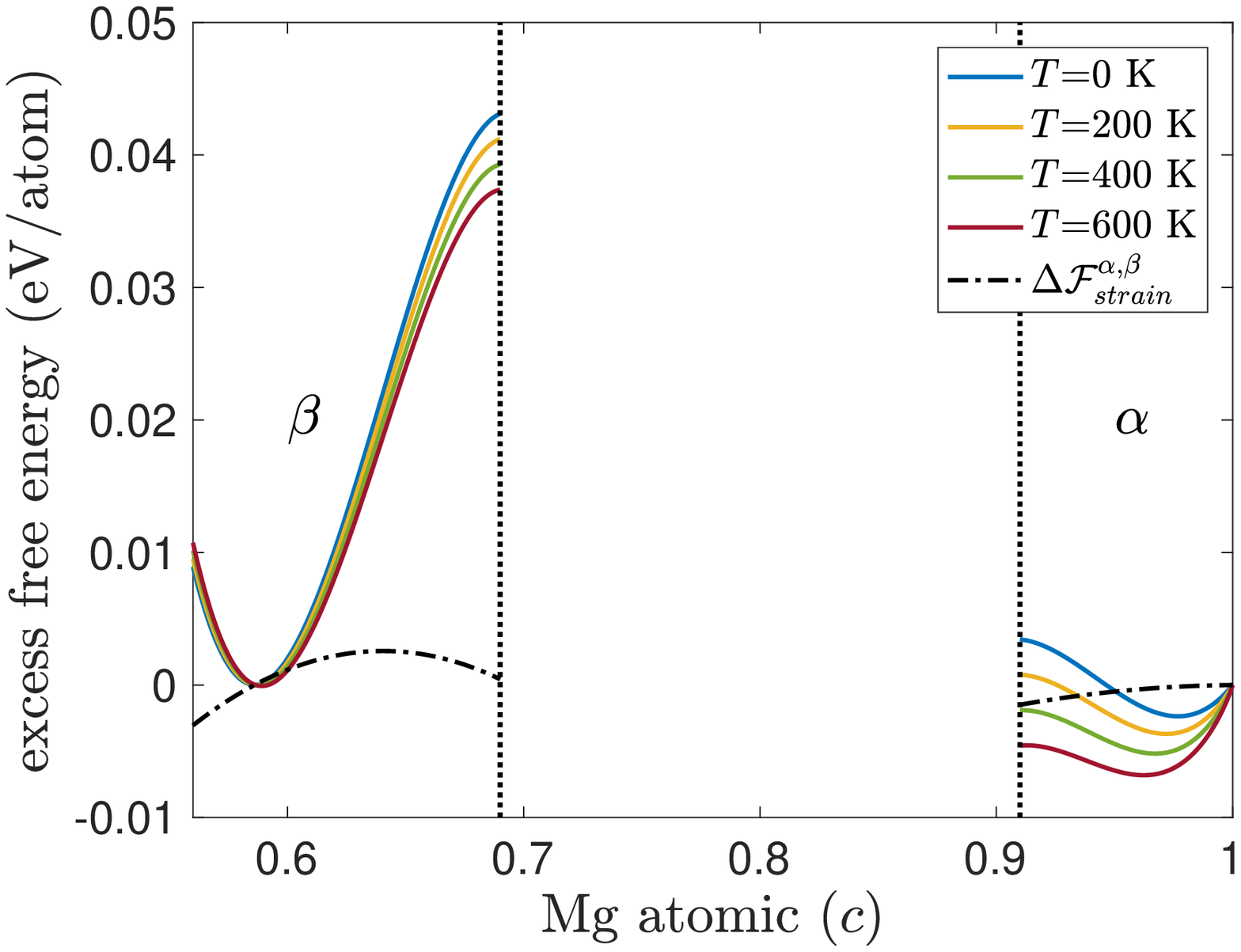}\label{Fig:FEUCb}}
\subfloat[uniaxial compression c (-0.03)]
{\includegraphics[keepaspectratio=true,width=0.35\textwidth]{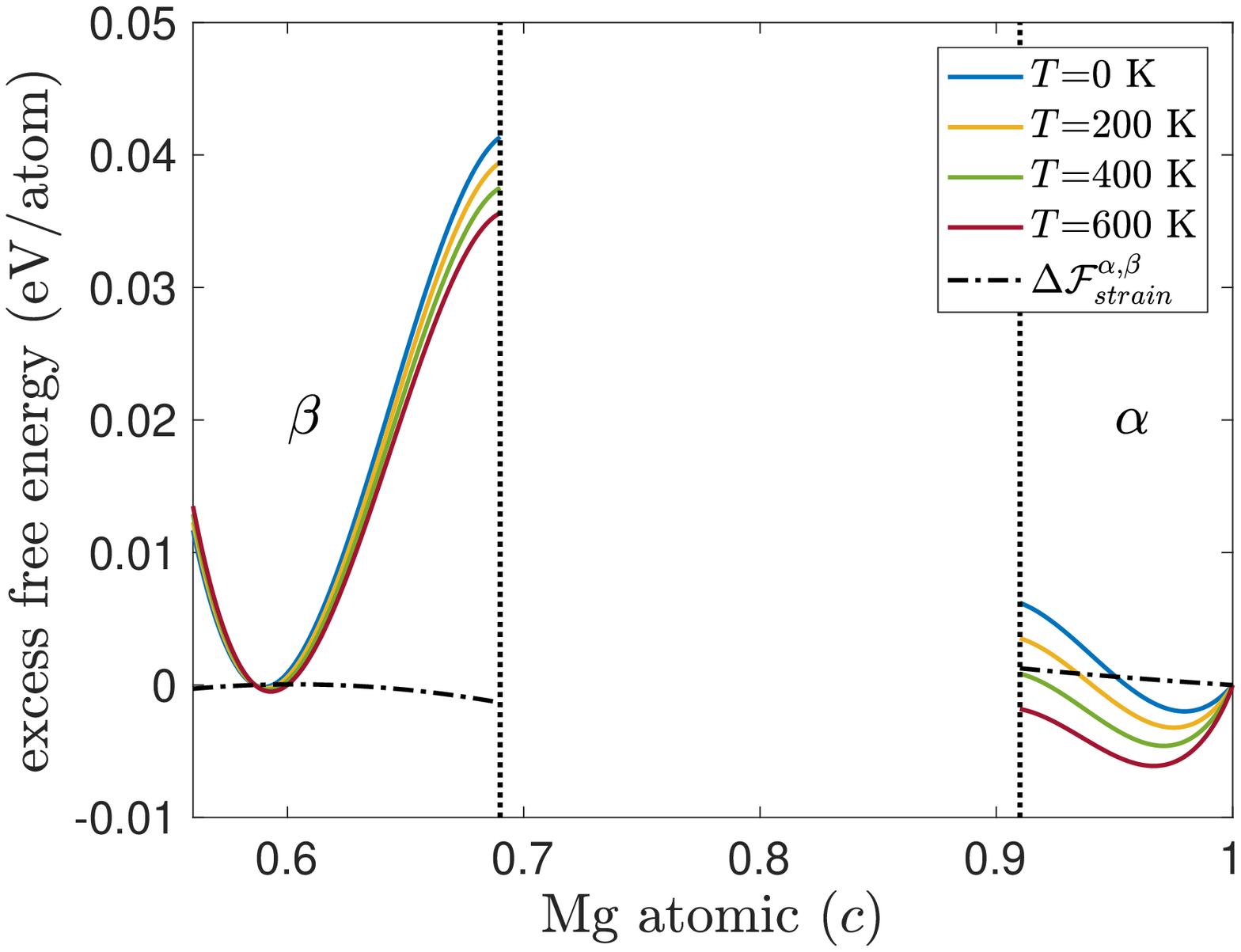}\label{Fig:FEUCc}}
\subfloat[volumetric tension (0.01)]
{\includegraphics[keepaspectratio=true,width=0.35\textwidth]{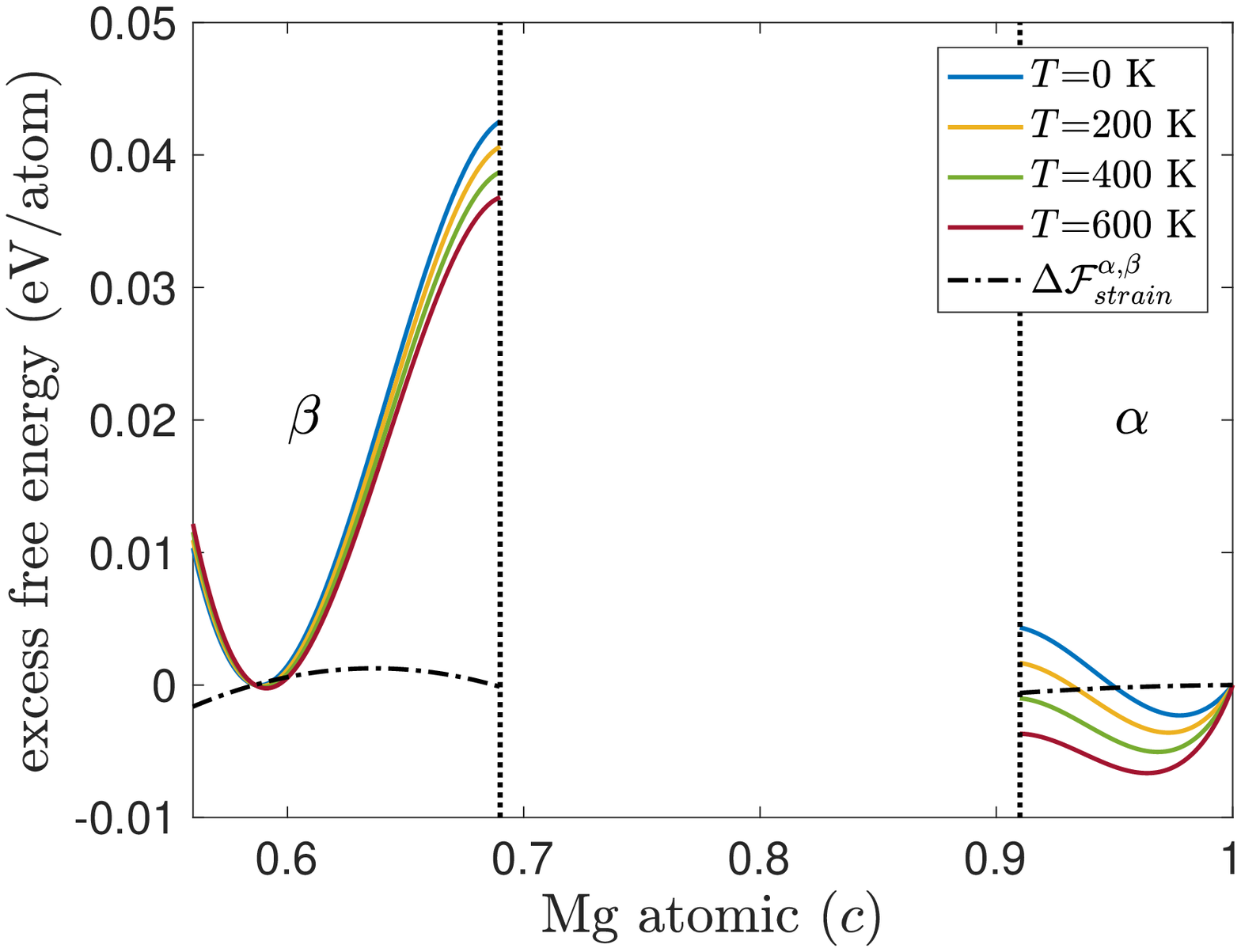}\label{Fig:FEVT}} \\
\subfloat[uniaxial tension a (0.03)] 
{\includegraphics[keepaspectratio=true,width=0.35\textwidth]{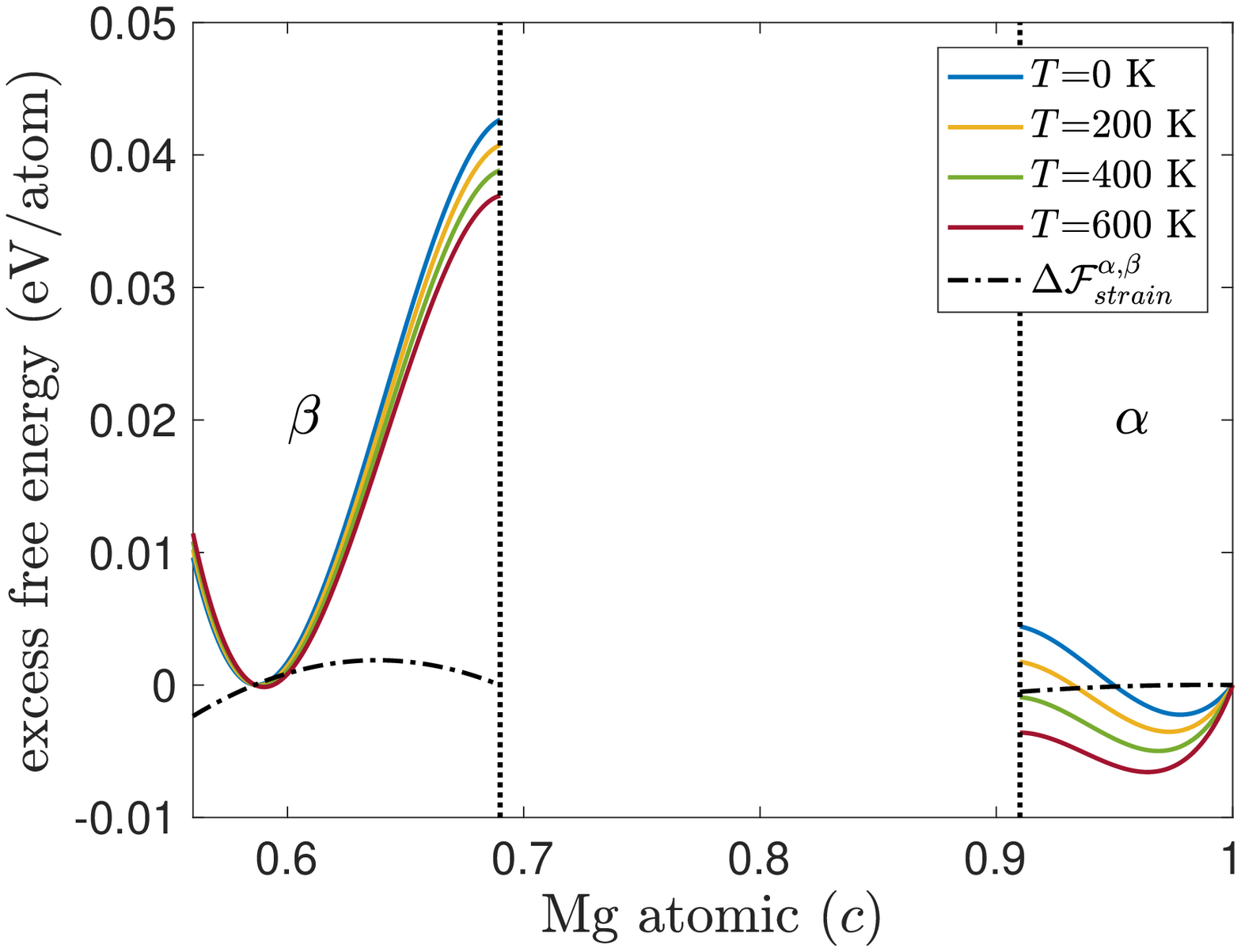}\label{Fig:FEUTa}}
\subfloat[uniaxial tension b (0.03)] 
{\includegraphics[keepaspectratio=true,width=0.35\textwidth]{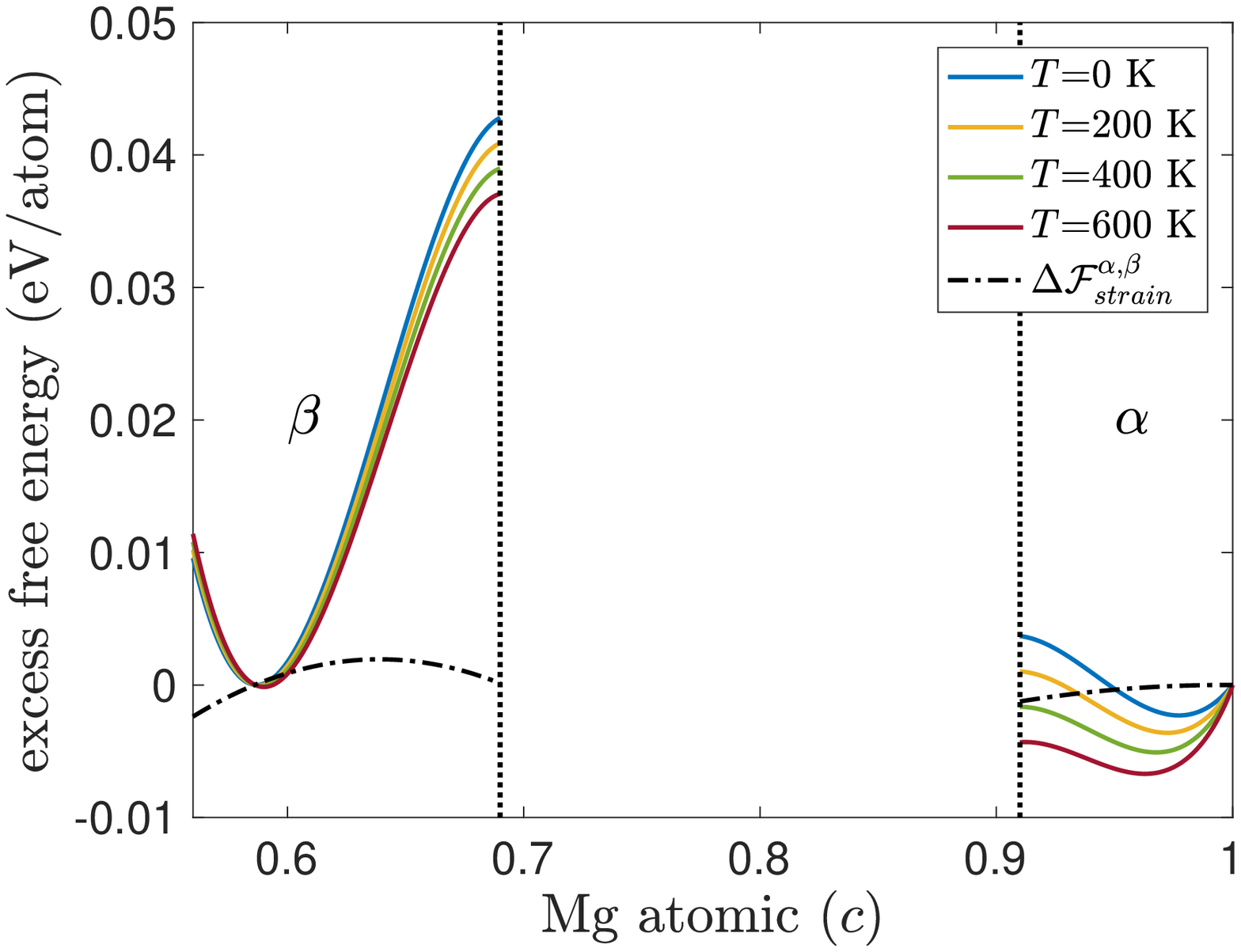}\label{Fig:FEUTb}}
\subfloat[uniaxial tension c (0.03)] 
{\includegraphics[keepaspectratio=true,width=0.35\textwidth]{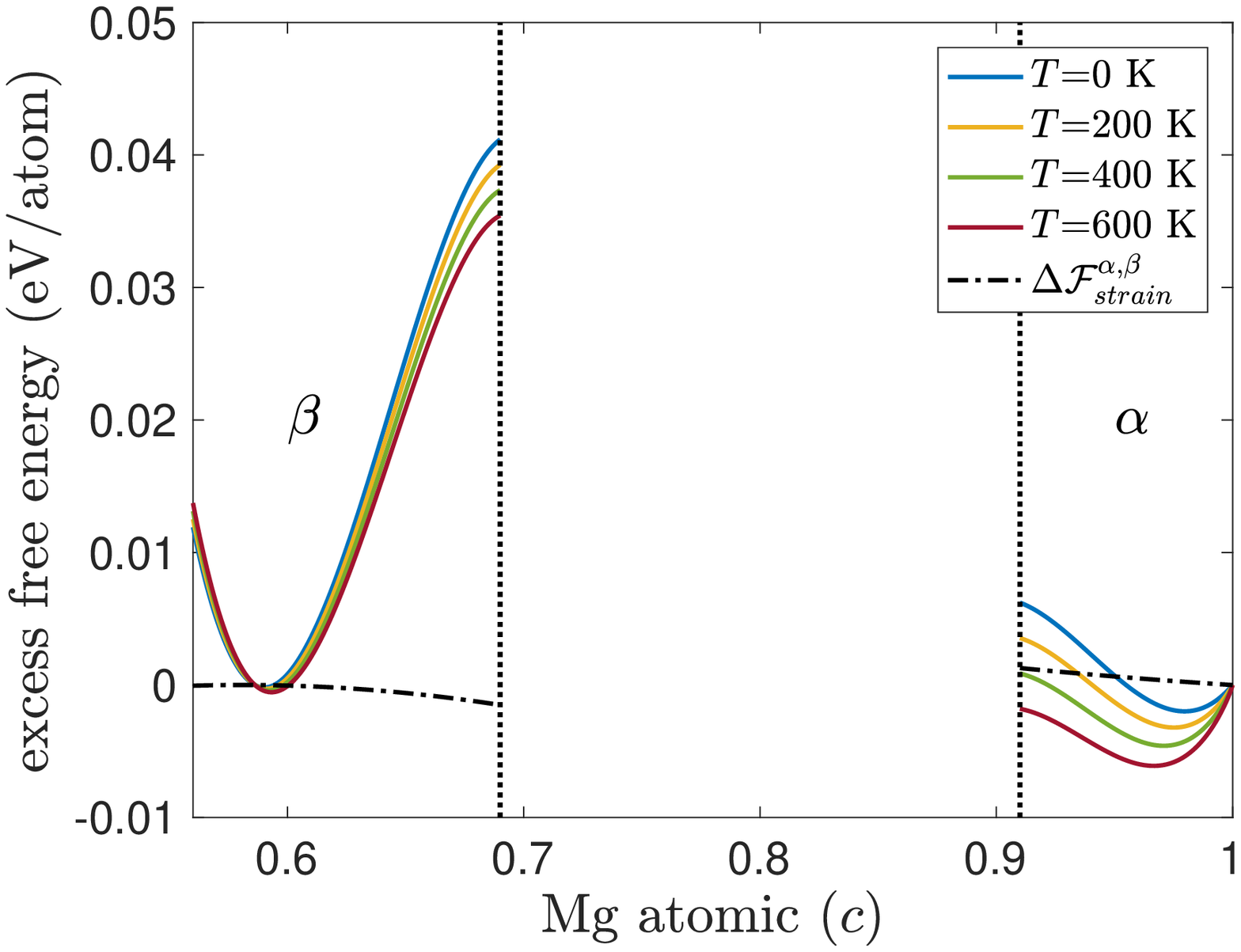}\label{Fig:FEUTc}} \\
\caption{Excess free energy of the solid solutions for 0, 200, 400 and 600 Kelvin temperatures under zero, tensile and compressive strains. } 
\end{figure}

We consider a case where the strains are $\be^{\alpha} = \be$, and $\be^{\beta}$ satisfies the Hadamard jump condition. The total strain in the $\alpha$ and $\beta$ phases are assumed to be independent of concentration. Fig \ref{Fig:FEZeroStrain} shows the concentration dependence of the excess free energy at zero strain, at temperatures 0, 200, 400 and 600 K. In Figs. \ref{Fig:FEVC} through \ref{Fig:FEUTc}, we plot the concentration dependence of the excess free energy for different cases of strains, at temperatures of 0, 200, 400 and 600 K. In these figures, we also plot the strain energy contribution ($\Delta \mathcal{F}^{\alpha,\beta}$) to the excess free energy as a function of concentration. Notice that $\Delta \mathcal{F}^{\alpha,\beta}$ not only depends on the concentration of each phase, but also on the nature of strain. From these figures, it is evident that the excess free energy for both $\beta$ and $\alpha$ phases are influenced by temperature and strain. For the $\beta$ phase solutions, the excess free energy is minimum at $c$ = $c_{\beta}$, and is positive elsewhere. For the $\alpha$ phase solutions, the excess free energy is zero at $c$ = 1, and has a minima between $0.9<c<1$. This minima shifts with temperature and the nature of strain. At $T = 0 K$, the excess free energy changes sign from negative to positive as $c$ is decreased from $1$ to $0.9$. As temperature is increased, the wells of excess free energy diagram for concentration range $0.9<c<1$ is lowered and the excess free energy between $0.9<c<1$ becomes negative. Overall, we conclude that thermomechanical loads have a considerable influence on the excess free energy in magnesium aluminum alloys.

\section{Concluding remarks}\label{Sec:Conclusion}
Summarizing, we have used first principles calculations to report the influence of thermomechanical loads on the energetics of precipitation in magnesium-aluminum alloys. Our DFT simulations show an increase in equilibrium volume (and hence the equilibrium lattice constants) with the increase in concentration of magnesium. Using these calculations, we have presented expressions of the energy of these solutions as a function of concentration, strain and temperature.  We have observed an increase in cohesive energies of solutions with increase in concentration, and also presented their dependence on strain.  We have also observed that the formation enthalpy of $\beta$ phase solutions to be strongly influenced by hydrostatic stress, however the formation enthalpy of $\alpha$ phase solutions remain unaffected by hydrostatic stress. Using the data from our first-principles calculations, we have presented an expression of the free energy of any magnesium aluminum solid solution, and takes into account the contributions of strain and temperature. We note that these expressions of the energy can serve as input to phase field simulations and process models of deformation driven precipitation in magnesium-aluminum alloys. 

It is typical during thermomechanical processing to encounter combined thermal and mechanical loadings. We have utilized these free energy expressions to report the dependence of equilibrium chemical potential, the solubility of magnesium in $\beta$ phase, and the solubility of aluminum in $\alpha$ phase on temperature and applied strain, which was heretofore unknown. We have showed that compressive strains along the $c$ axis decrease the solubility of Al in $\alpha$ phase, and tensile strains increase it. Strains along the $a$ and $b$ axes do not influence the solubility of Al in $\alpha$ phase. Increase in temperature also increases the solubility of Al in $\alpha$ phase. We have also reported a slight variation of the solubility of Mg in $\beta$ phase with temperature and strain. The equilibrium chemical potential is increased by applied compressive strains along the $c$ axis and increase in temperature, and it is decreased by tensile strains along the $c$ axis. Strains along the $a$ and $b$ axis do not influence the equilibrium chemical potential.

Finally, our calculations show that there is significant interplay between thermomechanical loads, and the growth of precipitates in magnesium-aluminum binary alloys. To this end, we have used the free energy surfaces to analyze the conditions of applied strains and temperature on the growth of precipitates. We have concluded that compressive strains along the $c$ axis direction leads to the growth of precipitates, whereas tensile strains along the $c$ axis impedes it's growth. Strains along $a$ and $b$ axis directions do not influence the growth of precipitates. Additionally, lowering of temperature promotes the thermodynamic driving force for precipitate growth. This insight can be used to control the final microstructure of the processed Mg-Al alloys. To elucidate, during thermomechanical processing such as rolling, the normal direction predominantly coincide with the $c$ axis. Additional strains applied along the normal direction - such as during confined rolling - can be used to control the growth of precipitates.

\section*{Acknowledgements}
The computations presented here were conducted on the Caltech High Performance Cluster partially supported by a grant from the Gordon and Betty Moore Foundation. Research was sponsored by the Army Research Laboratory and was accomplished under Cooperative Agreement Number W911NF-12-2-0022. The views and conclusions contained in this document are those of the authors and should not be interpreted as representing the official policies, either expressed or implied, of the Army Research Laboratory or the U.S. Government. The U.S. Government is authorized to reproduce and distribute reprints for Government purposes notwithstanding any copyright notation herein.

\bibliographystyle{unsrt}
\bibliography{gb_mgal}

\begin{thebibliography}{10}

\bibitem{Kainer:2000}
K.~U. Kainer and B.~L. Mordike.
\newblock {\em Magnesium alloys and their applications}.
\newblock Wiley-VCH Weinheim, Germany:, 2000.

\bibitem{Nie:2012}
J.-F. Nie.
\newblock Precipitation and hardening in magnesium alloys.
\newblock {\em Metallurgical and Materials Transactions A}, 43(11):3891--3939,
  Nov 2012.

\bibitem{Polmear:1994}
I.~J. Polmear.
\newblock Magnesium alloys and applications.
\newblock {\em Materials Science and Technology}, 10(1):1--16, 1994.

\bibitem{Nembach:1997}
E.~Nembach.
\newblock Particle strengthening of metals and alloys.
\newblock 1997.

\bibitem{Wilson:2003}
R.~Wilson, C.~J. Bettles, B.~C. Muddle, and J.~F. Nie.
\newblock Precipitation hardening in mg-3 wt\% nd (-zn) casting alloys.
\newblock In {\em Materials Science Forum}, volume 419, pages 267--272. Trans
  Tech Publications Ltd., Zurich-Uetikon, Switzerland, 2003.

\bibitem{He:2006}
Zeng X.~Q. He~S.~M., L.~M. Peng, X.~Gao, J.~F. Nie, and W.~J. Ding.
\newblock Precipitation in a mg-10gd-3y-0.4 zr (wt.\%) alloy during isothermal
  ageing at 250c.
\newblock {\em Journal of Alloys and Compounds}, 421(1/2):309--313, 2006.

\bibitem{Buha:2008}
J.~Buha and T.~Ohkubo.
\newblock Natural aging in mg-zn (-cu) alloys.
\newblock {\em Metallurgical and materials transactions A}, 39(9):2259, 2008.

\bibitem{Zhang:2018}
Y.~Zhang, W.~Rong, Y.~Wu, L.~Peng, J.-F. Nie, and N.~Birbilis.
\newblock A comparative study of the role of ag in microstructures and
  mechanical properties of mg-gd and mg-y alloys.
\newblock {\em Materials Science and Engineering: A}, 731:609--622, 2018.

\bibitem{Nakata:2017}
T.~Nakata, C.~Xu, R.~Ajima, K.~Shimizu, S.~Hanaki, T.~T. Sasaki, L.~Ma,
  K.~Hono, and S.~Kamado.
\newblock Strong and ductile age-hardening mg-al-ca-mn alloy that can be
  extruded as fast as aluminum alloys.
\newblock {\em Acta Materialia}, 130:261--270, 2017.

\bibitem{Nie:2003}
J.~F. Nie.
\newblock Effects of precipitate shape and orientation on dispersion
  strengthening in magnesium alloys.
\newblock {\em Scripta Materialia}, 48(8):1009--1015, 2003.

\bibitem{Ma:2019}
X.L. Ma, S.~E. Prameela, P.~Yi, M.~Fernandez, N.~M. Krywopusk, L.~J. Kecskes,
  T.~Sano, M.~L. Falk, and T.~P. Weihs.
\newblock Dynamic precipitation and recrystallization in mg-9wt.
  equal-channel angular extrusion: A comparative study to conventional aging.
\newblock {\em Acta Materialia}, 172:185 -- 199, 2019.

\bibitem{Prameela:2019}
S.~E. Prameela, P.~Yi, B.~Mediros, V.~Liu, L.~J. Keckes, M.~L. Falk, and T.~P.
  Weihs.
\newblock Deformation assisted nucleation of continuous nanoprecipitates in
  mg-al alloys.
\newblock {\em Available at SSRN 3484667}, 2019.

\bibitem{Lloyd:2019}
J.~T. Lloyd, A.~J. Matejunas, R.~Becker, T.~R. Walter, M.~.W Priddy, and
  J.~Kimberley.
\newblock Dynamic tensile failure of rolled magnesium: Simulations and
  experiments quantifying the role of texture and second-phase particles.
\newblock {\em International Journal of Plasticity}, 114:174--195, 2019.

\bibitem{Braszczynska:2009}
K.~N. Braszczy{\'n}ska-Malik.
\newblock Discontinuous and continuous precipitation in magnesium--aluminium
  type alloys.
\newblock {\em Journal of Alloys and Compounds}, 477(1-2):870--876, 2009.

\bibitem{Duly:1995}
D.~Duly, J.~P. Simon, and Y.~Brechet.
\newblock On the competition between continuous and discontinuous
  precipitations in binary mg-al alloys.
\newblock {\em Acta metallurgica et materialia}, 43(1):101--106, 1995.

\bibitem{Avedesian:1999}
M.~M. Avedesian, H.~Baker, et~al.
\newblock {\em ASM specialty handbook: magnesium and magnesium alloys}.
\newblock ASM international, 1999.

\bibitem{Caceres:2002}
C.~H. Caceres, C.~J. Davidson, J.~R. Griffiths, and C.~L. Newton.
\newblock Effects of solidification rate and ageing on the microstructure and
  mechanical properties of az91 alloy.
\newblock {\em Materials Science and Engineering: A}, 325(1-2):344--355, 2002.

\bibitem{Maitrejean:1999}
S.~Maitrejean, M.~Veron, Y.~Brechet, and G.~R. Purdy.
\newblock Morphological instabilities in mg-7.7 at\% al.
\newblock {\em Scripta materialia}, 41(11), 1999.

\bibitem{Nie:2001}
J.~F. Nie, X.~L. Xiao, C.~P. Luo, and B.~C. Muddle.
\newblock Characterisation of precipitate phases in magnesium alloys using
  electron microdiffraction.
\newblock {\em Micron}, 32(8):857--863, 2001.

\bibitem{Park:2012}
S.S. Park and B.S. You.
\newblock Low-temperature superplasticity of extruded mg--sn--al--zn alloy.
\newblock {\em Scripta Materialia}, 65(3):202 -- 205, 2011.

\bibitem{Zhang:2011}
J.~Zhang, Y.~Dou, B.~Zhang, and X.~Luo.
\newblock Elevated-temperature plasticity and mechanical properties of a rare
  earth-modified mg--zn--al alloy.
\newblock {\em Materials Letters}, 65(6):944 -- 947, 2011.

\bibitem{Mathis:2005}
K.~M{\'a}this, J.~Gubicza, and N.~H. Nam.
\newblock Microstructure and mechanical behavior of az91 mg alloy processed by
  equal channel angular pressing.
\newblock {\em Journal of Alloys and Compounds}, 394(1-2):194--199, 2005.

\bibitem{Ghosh:2017a}
S.~Ghosh and P.~Suryanarayana.
\newblock Sparc: Accurate and efficient finite-difference formulation and
  parallel implementation of density functional theory: Isolated clusters.
\newblock {\em Computer Physics Communications}, 212:189 -- 204, 2017.

\bibitem{Ghosh:2017b}
S.~Ghosh and P.~Suryanarayana.
\newblock Sparc: Accurate and efficient finite-difference formulation and
  parallel implementation of density functional theory: Extended systems.
\newblock {\em Computer Physics Communications}, 216:109 -- 125, 2017.

\bibitem{Perdew:1992}
J.~P. Perdew and Y.~Wang.
\newblock Accurate and simple analytic representation of the electron-gas
  correlation energy.
\newblock {\em Physical Review B}, 45(23):13244, 1992.

\bibitem{Ceperley:1980}
D.~M. Ceperley and B.~J. Alder.
\newblock Ground state of the electron gas by a stochastic method.
\newblock {\em Phys. Rev. Lett.}, 45:566--569, Aug 1980.

\bibitem{Troullier:1991}
N.~Troullier and J.~L. Martins.
\newblock Efficient pseudopotentials for plane-wave calculations.
\newblock {\em Physical Review B}, 43(3):1993--2006, 1991.

\bibitem{Banerjee:2016}
A.~S. Banerjee, P.~Suryanarayana, and J.~E. Pask.
\newblock Periodic pulay method for robust and efficient convergence
  acceleration of self-consistent field iterations.
\newblock {\em Chemical Physics Letters}, 647:31 -- 35, 2016.

\bibitem{Chou:1986}
M.Y. Chou and M.~L. Cohen.
\newblock Ab initio study of the structural properties of magnesium.
\newblock {\em Solid State Communications}, 57(10):785 -- 788, 1986.

\bibitem{Kittel}
C.~Kittel et~al.
\newblock {\em Introduction to solid state physics}, volume~8.
\newblock Wiley New York, 1976.

\bibitem{Koster:1961}
W.~Koster and H.~Franz.
\newblock Poisson's ratio for metals and alloys.
\newblock {\em Metallurgical Reviews}, 6(1):1--56, 1961.

\bibitem{Duan:2011}
Y.-H. Duan, Y.~Sun, M.-J. Peng, and Z.-Z. Guo.
\newblock First principles investigation of the binary intermetallics in
  pb--mg--al alloy: stability, elastic properties and electronic structure.
\newblock {\em Solid State Sciences}, 13(2):455--459, 2011.

\bibitem{Zhang:2005}
M-X Zhang and P.~M. Kelly.
\newblock Edge-to-edge matching and its applications: part ii. application to
  mg--al, mg--y and mg--mn alloys.
\newblock {\em Acta materialia}, 53(4):1085--1096, 2005.

\bibitem{Shin:2010}
D.~Shin and C.~Wolverton.
\newblock First-principles study of solute–vacancy binding in magnesium.
\newblock {\em Acta Materialia}, 58(2):531 -- 540, 2010.

\bibitem{Wang:2008}
N.~Wang, W.-Y. Yu, B.-Y. Tang, L.-M. Peng, and W.-J. Ding.
\newblock Structural and mechanical properties of mg 17 al 12 and mg 24 y 5
  from first-principles calculations.
\newblock {\em Journal of Physics D: Applied Physics}, 41(19):195408, 2008.

\bibitem{Han:2015}
G.~Han, Z.~Han, A.~A. Luo, and B.~Liu.
\newblock Three-dimensional phase-field simulation and experimental validation
  of $\beta$-mg 17 al 12 phase precipitation in mg-al-based alloys.
\newblock {\em Metallurgical and Materials Transactions A}, 46(2):948--962,
  2015.

\bibitem{Levenberg:1944}
K.~Levenberg.
\newblock A method for the solution of certain non-linear problems in least
  squares.
\newblock {\em Quarterly of applied mathematics}, 2(2):164--168, 1944.

\bibitem{Marquardt:1963}
D.~W. Marquardt.
\newblock An algorithm for least-squares estimation of nonlinear parameters.
\newblock {\em Journal of the society for Industrial and Applied Mathematics},
  11(2):431--441, 1963.

\end{thebibliography}
\end{document}